\documentclass[a4paper,fleqn,usenatbib]{mnras}

\usepackage[T1]{fontenc}
\usepackage{ae,aecompl}

\usepackage{graphicx}	
\usepackage{amsmath}	
\usepackage{amssymb}	
\usepackage{pdflscape}	
\usepackage{float}
\usepackage[usenames]{color}
\usepackage{textcomp}

\title[Clumpy torus model predictions in the mid-infrared]{A mid-infrared statistical investigation of clumpy torus model predictions}

\author[Garc\'ia-Gonz\'alez et
al.]{J. Garc\'ia-Gonz\'alez,$^1$\thanks{E-mail: jgarcia@ifca.unican.es}
  A. Alonso-Herrero,$^{2,3,4}$\thanks{E-mail: aalonso@cab.inta-csic.es} S. F. H\"onig,$^5$ A. Hern\'an-Caballero,$^6$
  \newauthor
  C. Ramos Almeida,$^{7,8}$ N. A. Levenson,$^{9,10}$ P.F. Roche,$^{3}$ O. Gonz\'alez-Mart\'{\i}n,$^{11}$
  \newauthor
C. Packham,$^{4,12}$ M. Kishimoto,$^{13}$   
\\
$^1$Instituto de F\'isica de Cantabria, CSIC-UC, Avenida de los Castros s/n, 39005 Santander, Spain\\
$^2$Centro de Astrobiolog\'ia (CSIC-INTA), ESAC Campus, E-28692 Villanueva de
la Ca\~nada, Madrid, Spain\\
$^{3}$Department of Physics, University of Oxford, Oxford OX1 3RH, UK\\
$^{4}$Physics \& Astronomy Department, University of Texas at San Antonio, San Antonio, Texas, 78249, USA\\ 
$^5$Department of Physics \& Astronomy, University of Southampton, Southampton SO17 1BJ, UK\\
$^{6}$Departamento de Astrof\'{\i}sica, Facultad de CC. F\'{\i}sicas, Universidad Complutense de Madrid, 28040 Madrid, Spain\\
$^7$Instituto de Astrof\'{\i}sica de Canarias (IAC), E-38205 La Laguna, Tenerife, Spain\\
$^8$Departamento de Astrof\'{\i}sica, Universidad de la Laguna (ULL), E-38206 La Laguna, Tenerife, Spain\\
$^9$Gemini Observatory, Casilla 603, La Serena, Chile\\
$^{10}$Space Telescope Science Institute, Baltimore, MD 21218, USA\\
$^{11}$Centro de Radioastronom\'{\i}a y Astrof\'{\i}sica (CRyA-UNAM), 3-72 (Xangari), 8701 Morelia, Mexico\\
$^{12}$National Astronomical Observatory of Japan, 2-21-1 Osawa, Mitaka, Tokyo 181-8588, Japan\\
$^{13}$Department of Astrophysics \& Atmospheric Sciences, Kyoto Sangyo University, Kamigamo-motoyama,\\
Kita-ku, Kyoto 605-8555, Japan 
}

\date{Accepted XXX. Received YYY; in original form ZZZ}

\pubyear{2017}

\begin{document}
\pagerange{\pageref{firstpage}--\pageref{lastpage}}
\maketitle

\begin{abstract}
We present new calculations of the CAT3D clumpy torus models, which now include 
  a more physical dust sublimation model as well as AGN anisotropic emission.
  These new models allow graphite grains to persist at temperatures higher than the silicate
  dust sublimation temperature. This produces stronger near-infrared emission
  and bluer mid-infrared (MIR) spectral slopes. 
We make a statistical comparison of the CAT3D model MIR
predictions with 
a compilation of sub-arcsecond resolution ground-based MIR spectroscopy of 52
nearby Seyfert galaxies (median distance of 36 Mpc) and 10 quasars. We focus on the 
AGN MIR spectral index $\alpha_{\rm MIR}$ and the strength of the $9.7\,\mu$m silicate feature $S_{\rm Sil}$.
As with other clumpy torus models, the new CAT3D models
do not reproduce the Seyfert galaxies with deep silicate
absorption ($S_{\rm Sil}<-1$).
Excluding those, we conclude that  the new CAT3D models are in better agreement with the observed
$\alpha_{\rm MIR}$ and $S_{\rm Sil}$ of
Seyfert galaxies and quasars.
We find that Seyfert 2 are reproduced with models with low photon escape 
probabilities, while the quasars and the Seyfert 1-1.5  require
  generally models with higher  photon escape probabilities. 
Quasars and Seyfert 1-1.5 tend to show steeper radial cloud
distributions and fewer clouds along an equatorial line-of-sight than Seyfert 2.
Introducing AGN anisotropic emission besides the more physical dust sublimation models
alleviates the problem of requiring inverted radial
  cloud distributions (i.e., more clouds towards the outer parts of the torus)
  to explain the MIR spectral indices of type 2 Seyferts.

\end{abstract}



\section{Introduction}

The dusty molecular torus is a key component of the Unification Model
\citep{Antonucci1993,Urry-Padovani1995} of Active Galactic Nuclei (AGN). Since
the torus was first proposed to explain the detection of polarised
broad lines in NGC~1068 and radio galaxies \citep{Antonucci1984} until its
first direct detection with ALMA in NGC1068
\citep{Garcia-Burillo2016,Gallimore2016}, our view of the torus has evolved enormously.

\begin{table*}
 \begin{center} 
 \caption{Summary of sample properties.}
 \label{galaxy-sample}
 \resizebox{13cm}{!}{
    \begin{tabular}{@{}lcccccc}
  \hline
      Name &     z &     $D _{\rm L}$   &      Morphological  &     $b/a$  &       Activity   &     Ref. Activity\\
  &  &      (Mpc)    &       type&&      type &    type \\
  \hline

    Circinus &   0.001448 &   4.2 &   SA(s)b? &   0.4 &   Sy 2      &   f\\
    ESO~103-G35 &    0.013286 &    59.1 &    S0? &    0.4 &    Sy 2  &   j, k, l\\
    ESO~323-G077 &    0.015014 &    60.2 &    (R)SAB0\textasciicircum 0(rs) &    0.7 &    Sy 1.2  &   k, l\\
    ESO~428-G14 &    0.005664 &    23.3 &    SAB0\textasciicircum 0(r) pec &    0.6 &    Sy 2  &     l\\
    IC~4329A &    0.016054 &    79.8 &    SA0\textasciicircum \ edge-on &    0.3 &    Sy 1.2  &   j, k, l\\
    IC~4518W &    0.016261 &    79.2 &    Sc pec &    0.5 &    Sy 2 &   k, l\\
    IC~5063 &    0.011348 &    49.9 &    SA0\textasciicircum +(s)? &    0.7 &    Sy 2  &    j, k\\
    MGC-3-34-64 &    0.016541 &    78.8 &    S0/a &    0.8 &    Sy 1.8  &    j, k\\
    MGC-5-23-16 &    0.008486 &    35.8 &    S0? &    0.5 &    Sy 2 &   j, k\\
    MCG-6-30-15 &    0.007749 &    26.8 &    S? &    0.6 &    Sy 1.2 &   j, k\\
    Mrk~3    &    0.013509 &    58.5 &    S0? &    0.9 &    Sy 2 &   j, k\\
    Mrk~1066 &    0.012025 &    49.0 &    (R)SB0\textasciicircum +(s) &    0.6 &    Sy 2  &      l\\
    Mrk~1210 &    0.013496 &    58.9 &    S? &    1.0 &    Sy 2 &      l\\
    Mrk~1239 &    0.019927 &    88.9 &    E-S0 &    1.0 &    Sy 1.5 &    h\\
    NGC~931  &    0.016652 &    67.5 &    SAbc &    0.2 &    Sy 1.5 &   j, k\\
    NGC~1068 &    0.003793 &    15.2 &    (R)SA(rs)b &    0.9 &    Sy 2 &   k\\
    NGC~1194 &    0.013596 &    54.5 &    SA0\textasciicircum +? &    0.6 &    Sy 1.9 &     l\\
    NGC~1320 &    0.008883 &    35.5 &    Sa? edge-on &    0.3 &    Sy 2      l\\
    NGC~1365 &    0.005457 &    21.5 &    SB(s)b &    0.6 &    Sy 1.5 &   i\\
    NGC~1386 &    0.002895 &    10.6 &    SB0\textasciicircum +(s) &    0.4 &    Sy 2 &    f\\
    NGC~1808 &    0.003319 &    12.3 &    (R)SAB(s)a &    0.6 &    Sy 2 &    b \\
    NGC~2110 &    0.007789 &    32.4 &    SAB0\textasciicircum - &    0.8 &    Sy 2 &    j, k\\
    NGC~2273 &    0.006138 &    28.7 &    SB(r)a? &    0.8 &    Sy 2 &   c \\
    NGC~2992 &    0.007710 &    34.4 &    Sa pec &    0.3 &    Sy 1.9 &     l\\
    NGC~3081 &    0.007976 &    34.5 &    (R)SAB0/a(r) &    0.8 &    Sy 2 &   j, k\\
    NGC~3094 &    0.008019 &    38.3 &    SB(s)a &    0.7 &    Sy 2 &   a \\
    NGC~3227 &    0.003859 &    20.4 &    SAB(s)a pec &    0.7 &    Sy 1.5 &   j, k, l\\
    NGC~3281 &    0.010674 &    45.0 &    SA(s)ab pec? &    0.5 &    Sy 2 &   j, k, l\\
    NGC~3783 &    0.009730 &    36.4 &    (R')SB(r)ab &    0.9 &    Sy 1 &   j, k\\
    NGC~4051 &    0.002336 &    12.9 &    SAB(rs)bc &    0.7 &    Sy 1.5 &    j, k\\
    NGC~4151 &    0.003319 &    20.0 &    (R')SAB(rs)ab? &    0.7 &    Sy 1.5  &    j, k, l\\
    NGC~4253 &    0.012929 &    61.3 &    (R')SB(s)a? &    0.8 &    Sy 1.5 &    j\\
    NGC~4258 &    0.001494 &    7.98 &    SAB(s)bc &    0.4 &    Sy 1.9 &   d\\
    NGC~4388 &    0.008419 &    17.0 &    SA(s)b? edge-on &    0.2 &    Sy 2 &   j, k\\
    NGC~4418 &    0.007268 &    34.9 &    (R')SAB(s)a &    0.5 &    Sy 2  &     l\\
    NGC~4507 &    0.011801 &    60.2 &    (R')SAB(rs)b &    0.8 &    Sy 2 &    j, k\\
    NGC~4579 &    0.00506  &    17.0 &    SAB(rs)bc &    0.8 &    Sy 1.9  &    d\\
    NGC~4593 &    0.009000 &    41.6 &    (R)SB(rs)b &    0.7 &    Sy 1  &   j, k, l\\
    NGC~5135 &    0.013693 &    58.3 &    SB(s)ab &    0.7 &    Sy 2 &      l\\
    NGC~5347 &    0.007789 &    40.2 &    (R')SB(rs)ab &    0.8 &    Sy 2 &      l\\
    NGC~5506 &    0.006181 &    30.1 &    Sa pec edge-on &    0.2 &    Sy 1.9 &   j, k\\
    NGC~5548 &    0.017175 &    80.3 &    (R')SA0/a(s) &    0.9 &    Sy 1.5 &   j, k, l\\
    NGC~5643 &    0.003999 &    14.4 &    SAB(rs)c &    0.9 &    Sy 2 &      l\\
    NGC~5995 &    0.025194 &    115  &    S(B)c &    0.9 &    Sy 2 &    k\\
    NGC~7130 &    0.016151 &    69.6 &    Sa pec &    0.9 &    Sy 2 &    f\\
    NGC~7172 &    0.008683 &    37.9 &    Sa pec edge-on &    0.6 &    Sy 2 &    j, k, l\\
    NGC~7213 &    0.005839 &    25.1 &    SA(s)a? &    0.9 &    Sy 1.5  &    j, k\\
    NGC~7465 &    0.006538 &    28.4 &    (R')SB0\textasciicircum 0?(s) &    0.7 &    Sy 2 &    e\\
    NGC~7469 &    0.016317 &    67.9 &    (R')SAB(rs)ab? &    0.7 &    Sy 1.2  &   j, k\\
    NGC~7479 &    0.007942 &    33.9 &    SB(s)c &    0.8 &    Sy 1.9 &     l\\
    NGC~7582 &    0.005254 &    22.1 &    (R')SB(s)ab &    0.4 &    Sy 2 &    j, k\\
    NGC~7674 &    0.028924 &    120  &    SA(r)bc pec &    0.9 &    Sy 2 &   g\\

  \hline
    
    \end{tabular}}
    \end{center}
    
    Notes.--- For each galaxy we give its redshift, luminosity distance ($D _{\rm L}$), morphological
    type, the axis ratio ($b/a$), and the  optical activity type.
    References.--- $^a$\citet{Asmus2014};  $^b$\citet{Brightman-Nandra};  $^c$\citet{Contini};
    $^d$\citet{Maiolino-Rieke}; $^e$\citet{Malizia}; $^f$\citet{Marinucci2012};
    $^g$\citet{Osterbrock-Martel}; $^h$\citet{Polletta1996}; $^i$\citet{Schulz1999};
    $^j$\citet{Tueller2008};
    $^k$\citet{Tueller2010}; $^l$\citet{Veron-CettyVeron2010}.

\end{table*}

\clearpage



From an observational point of view, high angular
resolution near and mid-infrared (NIR and MIR, respectively) imaging and spectroscopy
are now routinely isolating the unresolved emission believed to be
associated with dust heated by the AGN and in particular with dust in the torus
\citep{Mason2006,RamosAlmeida2009,Ramos-Almeida2011,AAH2011,Asmus2014,GarciaBernete2016}. The modelling of 
MIR interferometric observations infers a compact central obscuring source with a size typically
of less than approximately 10\,pc
\citep[MIR half-light radii, see][]{Tristram2009,Burtscher2013}. 
However, some recent  findings are starting to complicate this simple
scenario of the torus as an isolated structure.
For instance, some Seyfert galaxies show a large fraction of
the nuclear dust emission in the polar direction \citep{Hoenig2013,LopezGonzaga2016}. Also, it is clear that in
many AGN there are nuclear extended dust components (e.g., dust lanes, dust in the
narrow line region and/or ionisation cones)  not necessarily
associated with the dusty torus
\citep[see e.g.,][]{Radomski2003,Packham2005Circinus,Mason2006,Roche2006,Roche2007,AAH2011,Asmus2016,GarciaBernete2016}.

Torus models are also quickly improving. The early models had the 
dust uniformly distributed  \citep{PierKrolik1992,GranatoDanese1994,Stenholm1994,Efstathiou1995,Manske1998,vanBemmel2003,Schartmann2005,Fritz,Feltre2012}
mostly due to computational reasons. However, it was understood that the obscuring
material in the torus had to be in clouds because
otherwise it would be very difficult for the dust
to survive in this region \citep{Krolik1988}.
For more than a decade now, the majority of
torus models adopt a clumpy distribution for the dust
 \citep[see e.g.,][]{Rowan-Robinson1995,Nenkova2002,Nenkova2008a,Nenkova2008b,Dullemond2005,Hoenig2006,Schartmann2008,Hoenig-kishimoto,Stalevski2016}.
The newest models now include a more realistic representation of the torus by
having a clumpy two-phase medium \citep{Stalevski2012,Siebenmorgen2015}
 and a clumpy dusty torus with a polar outflow \citep{Hoenig2017}. Other models 
have radiation-driven obscuring structures which replace the {\it classical} torus \citep[see e.g.,][]{Wada2012}.

Clumpy torus models fit reasonably well the unresolved nuclear infrared 
emission of local AGN \citep{RamosAlmeida2009, Hoenig2010, Ramos-Almeida2011, AAH2011, Sales2011, Lira2013,
  Ichikawa2015, Siebenmorgen2015, Audibert2017}. However,
  in some cases fitting the entire nuclear infrared range simultaneously with clumpy torus models alone
  can be challenging
  \citep[see e.g.,][]{Mor2009,Garcia-Burillo2016, Martinez-Paredes2017}. Thus,
  there is still  room for improvement. In this work we
present new calculations of the \cite{Hoenig-kishimoto} torus models, known as CAT3D
models\footnote{http://www.sungrazer.org/cat3d.html}. We 
focus on two different improvements, namely, a more physical
description of the dust sublimation model and the inclusion of AGN anisotropic emission. Both
possibilities were discussed by \citet{Kishimoto2007} to explain 
the K-band interferometric observations of type 1 Seyferts.
As we shall see, this more physically  motivated dust sublimation
  model allows for the graphites grains to survive at higher temperatures than usually
  assumed for the silicates. This may alleviate the problem of the excess of nuclear NIR emission with
respect to the current clumpy torus predictions found mostly for type 1 
AGN \citep{Mor2009,AAH2011,Ichikawa2015} but also type 2 AGN \citep{Lira2013}. Indeed, \citet{Mor2009}
and \citet{Mor2012}
included a hot dust graphite component in addition to the {\it standard} torus
to model the infrared SEDs of type 1 AGN. 
Finally, with the availability of large samples of Seyfert galaxies
with sub-arcsecond resolution MIR spectroscopy
\citep{Hoenig2010,GonzalezMartin2013,AAH2016}, we can also compare the CAT3D torus old and
new model predictions
for the MIR nuclear emission with  AGN observations taking  a more statistical
approach.

This paper is organized as follows. Section~\ref{section_sample} describes the compilation of
MIR spectroscopy 
used in this work and Section~\ref{analysis} the MIR properties of the AGN emission as derived using a spectral
decomposition method. Section~\ref{models} presents the new CAT3D torus model calculations, as well as a comparison with
  the previous version  of the model and the MIR observations. We summarize our conclusions in
  Section~\ref{conclusions}.

\begin{figure*}

      \includegraphics[width=0.45\textwidth]{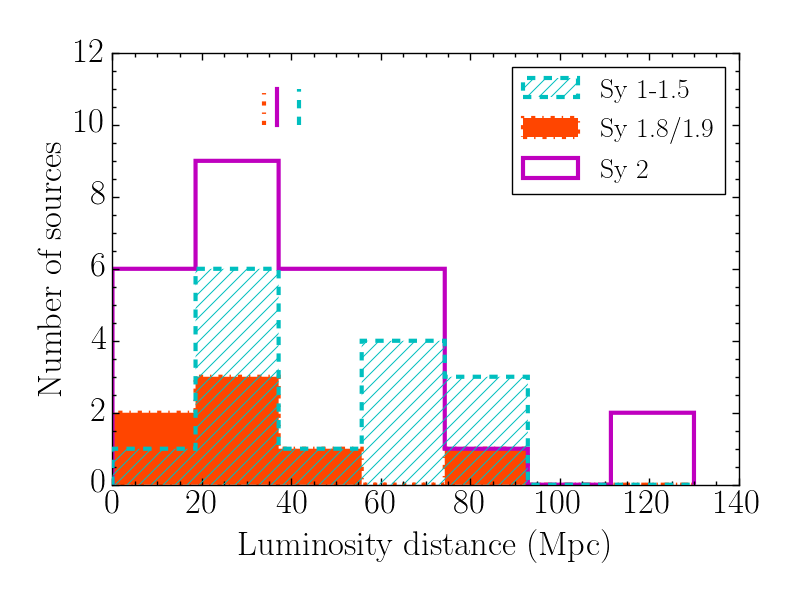}{\vspace{0cm}}
      \includegraphics[width=0.45\textwidth]{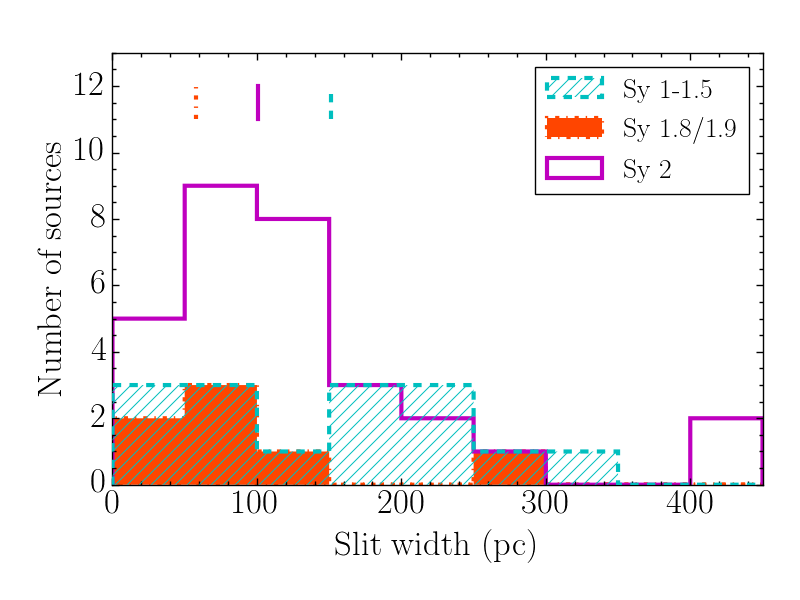}{\vspace{0cm}}

      \caption{Distribution of the luminosity distance (left)
        and the slit width (right) for the
        Seyfert 2 galaxies (30,  magenta histograms),
      Seyfert 1-1.5 galaxies (15, hatched cyan histograms),
      and  Seyfert 1.8/1.9 (7, filled orange histograms). The vertical lines indicate the
      corresponding median values of the distributions.}

  \label{fig_luminosity_distance}

\end{figure*}

\section{Sample and MIR ground based observations }
\label{section_sample}

We compiled a sample of 52 Seyfert galaxies
(see Table~\ref{galaxy-sample}) with existing high angular
resolution MIR spectroscopy (Table~\ref{table-instruments})
 obtained on $8-10\,$m class telescopes.
We chose instruments on large 
telescopes to take advantage of the angular
resolutions typically achieved in the MIR, $0.3-0.4$\,arcsec. This allows 
us to probe the nuclear regions of Seyfert galaxies with angular
resolutions almost a factor of ten better than with the {\it Spitzer} Infrared Spectrograph
\citep[IRS,][]{Houck2004}.
We used observations taken with four different instruments covering
the $N$-band atmospheric window, approximately between 7.5 and
13.5\,$\mu$m. The instruments include the Thermal-Region Camera Spectrograph
(T-ReCS; \citealt{Telesco1998}) on the 8.1\,m Gemini-South Telescope,
the Very Large Telescope (VLT) spectrometer and imager for the mid-infrared
(VISIR; \citealt{Lagage2004}) on the 8.2\,m VLT UT3 telescope 
at ESO/Paranal observatory,
the CanariCam instrument \citep{Telesco2003,Packham2005}
on the 10.4\,m Gran Telescopio CANARIAS (GTC) in El Roque de los Muchachos Observatory,
and Michelle \citep{Glasse1997} on the 8.1\,m Gemini-North Telescope.  
The CanariCam spectroscopy was taken as part of the ESO/GTC large programme 182.B-2005 (PI Alonso-Herrero).

Table~\ref{galaxy-sample} summarizes the properties of the Seyfert galaxies in our sample
including their redshift, luminosity distance ($H_0=73\,{\rm km\,s}^{-1}\,{\rm Mpc}
^{-1}$, $\Omega_M=0.27$ and $\Omega_\Lambda=0.73$), morphological
type, the axis ratio ($b/a$), and the  optical activity type (see
below).  We used the luminosity distance obtained
from the NASA Extragalactic Database (NED\footnote{http://ned.ipac.caltech.edu/})
using the corrected
redshift to the reference frame defined by the Virgo cluster, the Great
Attractor and the Shapley supercluster. We obtained the spectral
  classification of our galaxies from the literature (see last column of Table~\ref{galaxy-sample}
  for the references). We grouped together all Seyferts going from  Seyfert 1 to Seyfert 1.5 and those  
classified as Seyfert 1.8 and Seyfert 1.9.  This resulted in 15 galaxies in the Seyfert 1-1.5 group, 
7 in the Seyfert 1.8/1.9, and 30 in the Seyfert 2 group.
In Fig.~\ref{fig_luminosity_distance} (left panel)
we show the luminosity distance distributions for
the Seyfert 2, Seyfert 1.8/1.9, and  Seyfert 1-1.5.
The distances of the different type Seyfert galaxies are similar (median value of
36.1\,Mpc), although
the Seyfert 1-1.5 galaxies  are slightly further away (median of 42\,Mpc)
   than the Seyfert 2 galaxies  (median of 37\,Mpc) in our sample.

We obtained the ground-based high angular resolution MIR spectroscopy
from several works \citep[mostly from][but see 
  the last column of Table~\ref{table-instruments} for a
  complete list of references]{Hoenig2010,GonzalezMartin2013,Esquej2014,AAH2016}.
In Table~\ref{table-instruments} 
for each galaxy in the sample we summarize some of the
 observational details of the MIR spectroscopy, namely, the
 instrument,  the slit width (in arcsec and pc, respectively), and the reference to the published spectra. 
  Out of the  52
 galaxies, 17 were observed with CanariCam, 18 with T-ReCS, 23 with VISIR,
 and 1 with Michelle. 
  As can  be seen
 from this
 table,  7 galaxies in our
 sample were observed with two different instruments. For those
 cases, in Section~\ref{seccion-resultados-DeblendIRS} and the Appendix we 
 will discuss in detail which one is used for
 the analysis.
 As can also be seen
 from this table, the slit widths for the different instruments vary
 between 0.35\,arcsec and 0.75\,arcsec, which are appropriate for the
 image quality values (FWHM) of the observations in the MIR (typically $\le 0.4\,$arcsec, see
 e.g., \citealt{Hoenig2010} and \citealt{AAH2016}). For the distances of our
Seyfert galaxies, the slits probe nuclear regions between 7 and
 436 pc, with a
 median value of 101 pc for the entire sample.  Finally, we used
 for this work the fully reduced and calibrated 1-dimensional spectra of the galaxies.
 We refer the reader to the original articles (see Table~\ref{table-instruments}) for details
 of the observations and data reduction.
In Fig.~\ref{fig_luminosity_distance} (right panel) we show the slit width (in parsec) distributions for
 the Seyfert 2, Seyfert 1-1.5, and  Seyfert 1.8/1.9 in our sample.
The median of the slit width is
 larger for the Seyfert 1-1.5 galaxies (median physical sizes of
   151\,pc) than the Seyfert 2 (101\,pc), as expected because the former are more
   distant on average.


 Our sample of Seyfert galaxies is by necessity flux-limited in the MIR so
 high signal-to-noise ratio MIR spectroscopy could be obtained with reasonable integration
 times. Typically these MIR fluxes within
small apertures are above 20\,mJy and the acquisition of the target
also requires relatively compact morphologies (see for instance
\citealt{AAH2016} for a more detailed discussion). This results in typical
  AGN luminosities at rest-frame $12\,\mu$m   in the range
  $\log \nu L_\nu(12\mu{\rm m}) \sim 42-44\,{\rm erg \, s}^{-1}$
  \citep[see][and also Section~\ref{analysis}]{Hoenig2010, AAH2016}.  Although
not a complete sample, it is likely to be representative of the galaxies at the median
distance of the sample. For instance, it contains  80\% of the Seyferts in the complete
volume-limited sample (distances
between 10 and 40\,Mpc) selected from the nine-month Swift-BAT catalogue at $14-195\,$keV \citep{Tueller2008}
analysed by \citet{GarciaBernete2016}.


\begin{table}
\begin{center}
 \caption{Summary of MIR spectroscopic observations}
 \label{table-instruments}
 \resizebox{8cm}{!}{
    \begin{tabular}{@{}lcccc}
    
  \hline
      Name &    Instrument &    slit width (arcsec)  &     slit width (pc)  &    Ref  \\

  \hline

    Circinus &  {\bf VISIR}&  0.75&  15 &  d\\
     &  T-ReCS&  0.35&  7&  e, i \\
     ESO~103-G35 &    T-ReCS&  0.35&  100&  e \\
    ESO~323-G077 &   VISIR&  0.75&   219&  d\\
    ESO~428-G14 &    VISIR&  0.75&  85 &  d\\
    IC~4329A &    VISIR&  0.75&   290&  d\\
    IC~4518W &    T-ReCS&  0.70&  269&  e, f \\
    IC~5063 &    {\bf VISIR}&  0.75&  181 &  d\\
   &    T-ReCS&  0.65&  157&  e, j \\
    MGC-3-34-64 &    VISIR&  0.75&  287 &  d\\
    MGC-5-23-16 &   VISIR&  0.75&   130&  d\\
    MCG-6-30-15 &   VISIR&  0.75&   97&  d\\
    Mrk~3 &    CanariCam&  0.52&  147&  b\\
    Mrk~1066 &   CanariCam&  0.52&  124&  b\\
    Mrk~1210 &   CanariCam&  0.52&  148&  b\\
    Mrk~1239 &   VISIR&   0.75&   323&   c\\
    NGC~931 &    CanariCam&  0.52&  170&  b\\
    NGC~1068 &    VISIR&  0.4&   29&  d  \\
    NGC~1194 &    CanariCam&  0.52&  137&  b\\
    NGC~1320 &   CanariCam&  0.52&  89&  b\\
    NGC~1365 &    VISIR&   0.75&  78 &   c\\
  &    {\bf T-ReCS}&  0.35&  36&  e, g \\
    NGC~1386 &   T-ReCS&  0.31&  16&  e \\
    NGC~1808 &   T-ReCS&  0.35&  21&  e, k \\
    NGC~2110 &    VISIR&  0.75&   118&  d\\
    NGC~2273 &    CanariCam&  0.52&  75&  b\\
    NGC~2992 &  CanariCam&  0.52&  87&  b\\
    NGC~3081 &   T-ReCS&  0.65&  109&  e \\
    NGC~3094 &    T-ReCS&  0.35&  65&  e, h \\
    NGC~3227 &    VISIR&  0.75&  74 &  d\\
   &    {\bf CanariCam}&  0.52&  51&  b\\
    NGC~3281 &    {\bf VISIR}&   0.75&   164&   c\\
  &    T-ReCS&  0.35&  76&  e,l \\
    NGC~3783 &   VISIR&  0.75&  132 &  d\\
    NGC~4051 &   CanariCam&  0.52&  33&  b\\
    NGC~4151 &    Michelle&   0.36 &  35&  a\\
    NGC~4253 &   CanariCam&  0.52&  155&  b\\
    NGC~4258 &    CanariCam&  0.52&  20&  b\\
    NGC~4388 &    CanariCam&  0.52&  43&  b\\
    NGC~4418 &    T-ReCS&  0.35&  59&  e \\
    NGC~4507 &   VISIR&  0.75&   219&  d\\
    NGC~4579 &    CanariCam&  0.52&  43&  b\\
    NGC~4593 &    VISIR&  0.75&   151&  d\\
    NGC~5135 &    T-ReCS&  0.70&  198&  e, f \\
    NGC~5347 &   CanariCam&  0.52&  101&  b\\
    NGC~5506 &    T-ReCS&  0.35&  51&  e, h \\
    NGC~5548 &    CanariCam&  0.52&  202&  b\\
    NGC~5643 &    {\bf VISIR}&  0.75&   52&  d\\
   &    T-ReCS&  0.35&  24&  e \\
    NGC~5995 &    VISIR&  0.75&  418 &  d\\
    NGC~7130 &    T-ReCS&  0.70&  236&  e, f \\
    NGC~7172 &   T-ReCS&  0.35&  64&  e, h \\
    NGC~7213 &    VISIR&  0.75&   91&  d\\
    NGC~7465 &   CanariCam&  0.52&  72&  b\\
    NGC~7469 &   VISIR&  0.75&   247&  d\\
    NGC~7479 &    T-ReCS&  0.35&  58&  e \\
    NGC~7582 &    {\bf VISIR}&  0.75&  80 &  d\\
  &    T-ReCS&  0.70&  75&  e \\
    NGC~7674 &   VISIR&  0.75&   436&  d\\

  \hline

    \end{tabular}
 }
\end{center}

    Notes.--- For galaxies  observed with two different instruments,  bold-face indicate the one  
     used in Section~\ref{seccion-MIR-properties}
    (see Appendix for details).\\
     References:   
     $^a$\citet{AAH2011}; $^b$\citet{AAH2016}; $^c$\citet{Burtscher2013}; 
     $^d$\citet{Hoenig2010}; $^e$\citet{GonzalezMartin2013}; $^f$\citet{Diaz-Santos2010};
     $^g$\citet{AAH2012}; $^h$\citet{Roche2007};
      $^i$\citet{Roche2006}; $^j$\citet{Young2007}; $^k$\citet{Sales2013}; 
      $^l$\citet{Sales2011}.
\end{table}

\section{Spectral Decomposition}\label{analysis}

\subsection{The method}
\label{seccion-resultados-DeblendIRS}
The host galaxy may contribute a significant fraction of the MIR emission in Seyfert nuclei even at sub-arcsecond
        spatial resolution \citep[see][]{Hoenig2010,
          GonzalezMartin2013, Esquej2014, AAH2014, Asmus2014, AAH2016, GarciaBernete2016}. Since the MIR emission
        from dust heated by the AGN is considered unresolved even for the nearest galaxies in our sample,
        eliminating the host contribution allows a direct
        comparison of the nuclear spectra obtained with different physical apertures.

We use the \textsc{deblend}IRS tool\footnote{http://www.denebola.org/ahc/deblendIRS/}
 \citep{Antonio2015} to do the decomposition of the spectra. \textsc{deblend}IRS is an
 IDL/GDL routine that decomposes the MIR spectra in three
 components: AGN, stellar emission (STR), and 
 interstellar emission (PAH), using a large library of \textit{Spitzer}/IRS spectra as templates for these components.
 The \textsc{deblend}IRS  templates 
 representing the AGN, stellar and PAH emission spectra are derived from
 observations probing physical scales larger than those of our ground-based nuclear spectra.
When we perform the spectral decomposition of the nuclear spectra of
Seyfert galaxies, we are implicitly assuming that the PAH and stellar
emission in the MIR probed on kpc scales by the {\it Spitzer}/IRS spectra are also
representative of those on tens of parsecs. 
As shown by
\cite{Alonso-Herrero2016b}, the IRS templates work well for the
majority of nuclear regions hosting an AGN in local (U)LIRGs and quasars.
This probably indicates that the MIR 
emission associated with stars and the interstellar medium (ISM) is not fundamentally affected by the presence 
of the radiation field of the AGN, at least for the typical 
physical scales probed by the slits, 100-150pc \citep{Esquej2014,AAH2014}.
The only exception was for nuclei with deep silicate features.  As explained by 
\cite{Antonio2015}, this is because of the small number of AGN templates with deep silicate absorption
  ({\it obscured} templates). This is a consequence of the relative scarcity of IRS spectra for AGN that
  feature both deep silicate absorption and no PAH emission.

\begin{table*}
 \caption{\textsc{deblend}IRS results for the Seyfert galaxy sample. }
 \label{table-deblendIRS_results}
\begin{center}
\begin{tabular}{@{}lcccccccccccccc}
\hline
 Name & $\chi^2$ & AGN $\nu L_{\nu}$ (12\,$\mu$m) &\multicolumn{3}{c}{MIR Contribution} & AGN Frac. at 12\,$\mu$m& AGN $S_{\rm Sil}$ & AGN $\alpha_{\rm MIR}$ \\
  && (erg s$^{-1}$)& AGN & PAH & STR\\
\hline
\multicolumn{9}{c}{Seyfert 1-1.5 galaxies}\\
\hline
  ESO~323-G77 & 0.50 & $3.2 \times 10^{43}$ & 0.66 & 0.01 & 0.33 & 0.71 [0.53,  0.90] & -0.2 [-0.6, 0.2] & -1.9 [-2.7, -1.1]\\
  IC~4329A & 0.30 & $1.7 \times 10^{44}$ & 0.75 & 0.00 & 0.25 & 0.83  [0.72, 0.93] & -0.1  [-0.3, 0.1] & -2.0 [-2.5,  -1.5]\\
  MCG-6-30-15 & 0.49 & $5.8 \times 10^{42}$ & 0.80 & 0.03 & 0.17 & 0.82 [0.67, 0.94] & 0.0 [-0.3, 0.3] & -1.8 [-2.6, -1.4]\\
  Mrk~1239 & 0.25 & $7.3 \times 10^{43}$ & 0.78 & 0.01 & 0.21 & 0.86 [0.71, 0.94] & 0.2 [0.0, 0.3] & -1.7 [-2.6, -1.3]\\
  NGC~931 & 1.69 & $6.0 \times 10^{43}$ & 0.93 & 0.01 & 0.06 & 0.90 [0.78, 0.97] & -0.1 [-0.3, 0.2] & -2.0 [-2.5, -1.7]\\
  NGC~1365 (T)& 3.94 & $4.5 \times 10^{42}$ & 0.97 & 0.03 & 0.00 & 0.86 [0.81, 0.98] & 0.1 [-0.0, 0.3] & -2.3 [-2.9, -2.0]\\
  NGC~3227 (C)& 2.15 & $5.0 \times 10^{42}$ & 0.71 & 0.06 & 0.23 & 0.88 [0.81, 0.96] & -0.1 [-0.3, 0.2] & -2.4 [-2.8, -2.0]\\
  NGC~3783 & 1.89 & $2.0 \times 10^{43}$ & 1.00 & 0.00 & 0.00 & 0.86 [0.81, 0.98] & 0.0 [-0.2, 0.2] & -2.2  [-2.8, -1.9]\\
  NGC~4051 & 3.17 & $1.8 \times 10^{42}$ & 0.90 & 0.10 & 0.00 & 0.85 [0.74, 0.94] & 0.1 [-0.2, 0.3] & -2.1 [-2.8, -1.8]\\
  NGC~4151 & 1.73 & $1.5 \times 10^{43}$& 0.83 & 0.00 & 0.17 & 0.90 [0.84, 0.97] & -0.0 [-0.2, 0.2] & -2.3 [-2.8, -1.9]\\
  NGC~4253 & 1.94 & $3.7 \times 10^{43}$ & 0.91 & 0.03 & 0.06 & 0.92 [0.83, 0.97] & -0.2 [-0.4, 0.1] & -2.4 [-2.7, -2.1]\\
  NGC~4593 & 1.30 & $1.3 \times 10^{43}$ & 0.65 & 0.00 & 0.35 & 0.83 [0.68, 0.94] & 0.3 [0.1, 0.5] & -1.9 [-2.8, -1.3]\\
  NGC~5548 & 4.22 & $3.8 \times 10^{43}$ & 0.86 & 0.04 & 0.10 & 0.83 [0.69, 0.94] & 0.1 [-0.2, 0.4] & -1.7 [-2.5, -1.3]\\
  NGC~7213* & 3.89 & $4.4 \times 10^{42}$ & 1.00 & 0.00 & 0.00 & 0.99 [0.98, 1.00] & 0.5 [0.3, 0.6] & -2.2 [-2.3, -2.0]\\
  NGC~7469 & 1.54 & $7.5 \times 10^{43}$ & 0.99 & 0.00 & 0.01 & 0.85 [0.79, 0.97] & 0.1 [-0.2, 0.3] & -2.2 [-2.8, -1.8]\\
  
\hline
\multicolumn{9}{c}{Seyfert 1.8/1.9 galaxies}\\
\hline

  MCG-3-34-64 & 2.37 & $1.4 \times 10^{44}$ & 0.79 & 0.04 & 0.17 & 0.88 [0.79, 0.95] & -0.2 [-0.5, 0.0] & -2.3 [-2.7, -2.0]\\
  NGC~1194 & 1.59 & $1.5 \times 10^{43}$ & 0.78 & 0.00 & 0.22 & 0.87 [0.79, 0.91] & -1.2 [-1.4, -0.9] & -1.2 [-1.7, -1.0]\\
  NGC~2992 & 5.12 & $8.8 \times 10^{42}$ & 0.99 & 0.00 & 0.01 & 0.98 [0.96, 1.00] & -0.3 [-0.5, -0.2] & -2.7 [-2.9, -2.6]\\
  NGC~4258 & 3.11 & $2.6 \times 10^{41}$ & 0.72 & 0.00 & 0.28 & 0.89 [0.86, 0.93] & 0.3 [0.1, 0.4] & -2.7 [-2.9, -1.8]\\
  NGC~4579 & 5.13 & $7.3 \times 10^{41}$ & 0.87 & 0.00 & 0.13 & 0.96 [0.93, 0.98] & 0.4 [0.3, 0.6] & -2.1 [-2.4, -1.7]\\
  NGC~5506 & 0.30 & $2.7 \times 10^{43}$ & 0.80 & 0.05 & 0.15 & 0.68 [0.32,  0.90] & -1.1 [-2.6, -0.2] & -1.8 [-2.7, -0.9]\\
  NGC~7479 & 11.2 & $1.4 \times 10^{43}$ & 0.86 & 0.06 & 0.08 & 0.89 [0.83, 0.93] & -3.4 [-3.6, -2.7] & -1.6 [-1.8, -1.2]\\

\hline
\multicolumn{9}{c}{Seyfert 2 galaxies}\\
\hline

  Circinus (V)& 7.51 & $6.5 \times 10^{42}$ & 0.99 & 0.00 & 0.01 & 0.99 [0.97, 1.00] & -1.4 [-1.5, -1.2] & -1.9 [-2.0, -1.7]\\
  ESO~103-G35* & 1.54 & $5.7 \times 10^{43}$ & 0.97 & 0.03 & 0.00 & 0.97 [0.88, 0.99] & -0.8 [-1.0, -0.6] & -2.2 [-2.6, -1.9]\\
  ESO~428-G14 & 3.37 & $3.8 \times 10^{42}$ & 0.87& 0.13 & 0.00 & 0.90 [0.84, 0.96] & -0.6 [-0.8, -0.4] & -2.6 [-2.9,  -2.3]\\
  IC~4518W & 2.19 & $2.9 \times 10^{43}$ & 0.99 & 0.01 & 0.00 & 0.94 [0.83, 0.98] & -1.5 [-1.9, -1.2] & -2.0 [-2.4, -1.5]\\
  IC~5063 (V)& 2.18 & $7.5 \times 10^{43}$ & 0.82 & 0.00 & 0.18 & 0.93 [0.91,  0.97] & -0.3 [-0.5, -0.2] & -2.6 [-2.8, -2.2]\\
  MCG-5-23-16 & 1.39 & $2.5 \times 10^{43}$ & 0.93 & 0.02 & 0.05 & 0.89 [0.83, 0.96] & -0.4 [-0.5, -0.2] & -2.5 [-2.8, -2.1]\\
  Mrk~3 & 4.11 & $4.0 \times 10^{43}$ & 0.85 & 0.00 & 0.15 & 0.96 [0.93,  0.99] & -0.5 [-0.7, -0.3] & -2.8 [-3.0, -2.6]\\
  Mrk~1066 & 6.22 & $9.7 \times 10^{42}$ & 0.69 & 0.31 & 0.00 & 0.73 [0.62, 0.82] & -0.8 [-1.2, -0.6] & -2.6 [-3.0, -2.2]\\
  Mrk~1210 & 5.39 & $5.5 \times 10^{43}$ & 0.94 & 0.00 & 0.06 & 0.98 [0.96, 0.99] & -0.3 [-0.5, -0.2] & -2.7 [-2.9, -2.5]\\
  NGC~1068 & 1.00 & $8.0 \times 10^{43}$ & 0.80 & 0.00 & 0.20 & 0.87 [0.79, 0.92] & -0.4 [-0.6, -0.2] & -2.1 [-2.7, -1.9]\\
  NGC~1320 & 2.14 & $1.3 \times 10^{43}$ & 0.90 & 0.00 & 0.10 & 0.93 [0.85, 0.98] & -0.2 [-0.4, -0.0] & -2.4 [-2.7, -2.0]\\
  NGC~1386 & 1.33 & $7.9 \times 10^{41}$ & 0.82 & 0.07 & 0.11 & 0.85 [0.73, 0.93] & -0.8 [-1.2, -0.5] & -2.2 [-2.7, -1.7]\\
  NGC~1808 & 24.6 & $1.3 \times 10^{42}$ & 0.87 & 0.13 & 0.00 & 0.91 [0.87, 0.94] & -0.6 [-0.7, -0.4] & -2.9 [-3.0, -2.7]\\
  NGC~2110 & 2.11 & $8.4 \times 10^{42}$ & 0.81 & 0.06 & 0.13 & 0.88 [0.77, 0.95] & 0.2 [-0.0, 0.4] & -1.8 [-2.7, -1.5]\\
  NGC~2273 & 1.76 & $7.7 \times 10^{42}$& 0.93 & 0.00 & 0.07 & 0.97 [0.96, 0.99] & -0.4 [-0.5,  -0.2] & -2.7 [-2.9, -2.6]\\
  NGC~3081 & 0.93 & $5.9 \times 10^{42}$ & 1.00 & 0.00 & 0.00 & 0.91 [0.85, 0.97] & -0.1 [-0.3, 0.2] & -2.4 [-2.8, -2.0]\\
  NGC~3094 & 209 & $2.8 \times 10^{43}$ & 1.00 & 0.00 & 0.00 & 0.99 [0.98, 1.00] & -4.0 [-4.1, -3.8] & -0.5 [-0.6, -0.3]\\
  NGC~3281* (V)& 3.39 & $1.9 \times 10^{43}$ & 0.94 & 0.02 & 0.04 & 0.97 [0.93, 0.99] & -1.2 [-1.4, -1.1] & -1.4 [-1.6, -1.1]\\
  NGC~4388 & 11.5 & $2.6 \times 10^{42}$ & 0.69 & 0.20 & 0.11 & 0.85 [0.80, 0.95] & -1.1 [-1.4, -0.8] & -3.5 [-3.8, -2.5]\\
  NGC~4418* & 267 & $2.1 \times 10^{43}$ & 1.00 & 0.00 & 0.00 & 0.99 [0.98, 1.00] & -4.1 [-4.2, -3.9] & -1.8 [-2.0, -1.7]\\
  NGC~4507 & 0.81 & $5.5 \times 10^{43}$ & 0.80 & 0.00 & 0.20 & 0.84 [0.72, 0.95] & -0.0 [-0.3, 0.2] & -2.0 [-2.6, -1.5]\\
  NGC~5135 & 4.10 & $1.1 \times 10^{43}$ & 0.81 & 0.03 & 0.16 & 0.88 [0.80, 0.94] & -0.7 [-0.9, -0.5] & -2.4 [-2.7, -2.0]\\
  NGC~5347 & 6.83 & $1.5 \times 10^{43}$ & 1.00 & 0.00 & 0.00 & 0.96 [0.93, 0.99] & -0.3 [-0.4, -0.1] & -2.5 [-2.8, -2.3]\\
  NGC~5643 (V)& 6.65 & $1.7 \times 10^{42}$ & 0.95 & 0.05 & 0.00 & 0.97 [0.95, 0.99] & -0.5 [-0.7, -0.3] & -2.7 [-2.9, -2.6]\\
  NGC~5995 & 0.92 & $1.0 \times 10^{44}$ & 0.56& 0.03 & 0.41 & 0.77 [0.64, 0.90] & -0.3 [-0.6, -0.0] & -2.1 [-2.7,  -1.4]\\
  NGC~7130 & 1.49 & $1.9 \times 10^{43}$ & 0.83 & 0.17 & 0.00 & 0.89 [0.83, 0.95] & -0.6 [-0.8, -0.4] & -2.7 [-3.0, -2.4]\\
  NGC~7172 & 6.70 & $6.1 \times 10^{42}$ & 0.94 & 0.06 & 0.00 & 0.94 [0.80, 0.98] & -2.4 [-2.7, -1.8] & -1.0 [-1.9, -0.8]\\
  NGC~7465 & 2.33 & $1.4 \times 10^{42}$ & 0.67 & 0.11 & 0.22 & 0.77 [0.66, 0.89] & -0.1 [-0.4,  0.3] & -2.2 [-2.8, -1.7]\\
  NGC~7582 (V)& 3.29 & $6.6 \times 10^{42}$ & 0.89 & 0.11 & 0.00 & 0.86 [0.78, 0.94] & -1.1 [-1.4, -0.9] & -1.7 [-2.0, -1.3]\\
  NGC~7674 & 1.61 & $1.6 \times 10^{44}$ & 0.71 & 0.04 & 0.25 & 0.84 [0.76, 0.92] & -0.2 [-0.4, 0.1] & -2.2 [-2.7, -1.8]\\
\hline
\end{tabular}
\end{center}

Notes.---  The $\chi^2$ values are reduced ones.  The MIR
contributions of the AGN, PAH and STR components are estimated in the $5-15\,\mu$m range. 
The AGN MIR spectral index $\alpha_{\rm MIR}$ is estimated in the $8.1-12.5\,\mu$m range.  We give the median value and in
parenthesis the 16\% and 84\% 
percentiles of the distributions for the AGN fractional contribution
at 12\,$\mu$m (within the slit), the strength of the $9.7\,\mu$m silicate feature and the spectral index.
 The galaxies fitted with themselves are marked with an asterisk.

\end{table*}

In Table~\ref{table-deblendIRS_results} we list the results for the \textsc{deblend}IRS 
spectral decomposition for the galaxies of our sample. For each galaxy we provide
here the quantities relevant
to this study. These are the reduced $\chi^2$ value of the best-fit model, the rest-frame 12\,$\mu$m monochromatic AGN
luminosity, calculated using the best-fit AGN component
at that wavelength,  the best fit value of the AGN, PAH and STR
fractional contributions in the $5-15\,\mu$m range, the median value of the AGN fractional contribution 
  within the slit at
rest-frame 12\,$\mu$m, the median value of the AGN strength of the
$9.7\,\mu$m silicate feature $S_{\rm Sil}$ (positive values are for the
  feature in emission and negative for the feature in absorption), and the median value of the AGN
MIR spectral index in the $8.1-12.5\,\mu$m spectral range, $\alpha_{\rm MIR}$.
\textsc{deblend}IRS computes full probability distribution functions
(PDF) using a marginalisation method to provide  reliable expectation values and uncertainties for AGN
properties \citep[see][for full details]{Antonio2015}. Thus for the last three columns in
Table~\ref{table-deblendIRS_results}  we also list  
the 1$\sigma$ confidence interval (i.e., the 16\% and
84\% percentiles of the PDF). In the Appendix we show two examples of 
the \textsc{deblend}IRS graphical outputs and also discuss the cases of galaxies observed with two
different instruments.

 Among the cases with small reduced $\chi^2$ values,
there are 8 of them with $\chi^2 <1$, which can indicate correlated errors.
Most of the galaxies   (7 of 8) with reduced $\chi^2 <1$ correspond to VISIR
  spectra. The errors of the VISIR spectra include an additional correlated source of uncertainty which comes
  from the averaging
  (and deviation) of the chop-nod beams in each spectral setting. This is a significant component in the
  error budget which some times can even dominate the computed total error of the spectra.
  NGC~3094 and NGC~4418, have very large values of $\chi^2$   and large residuals, what means
  the fit is bad, due to their deep silicate absorption. 
As explained in \cite{Alonso-Herrero2016b}, 
for the deepest silicate feature the $\chi^2$ values worsen. 
\cite{Roche2015} compared the IRS and T-ReCS spectra of NGC~4418
and  found that T-ReCS spectrum only shows  
the deep silicate absorption whereas the larger IRS aperture spectrum shows  additionally other
spectral features that may be due to the diffuse emission of the host galaxy.
This causes differences in the spectral shape, which are not captured by the IRS templates, explaining the
high value of $\chi^2$. 
The same happens with NGC~3094, which was studied by
\cite{Roche2007}. They found
evidence  of spectral structure at 11\,$\mu$m that may explain 
the differences in shape between the T-ReCS spectrum and the IRS larger aperture spectrum. 
The shortage of templates with strong silicate absorption compared 
to the others also increases the value of $\chi^2$.


\begin{figure}
    \begin{center}
      
      \includegraphics[width=0.45\textwidth]{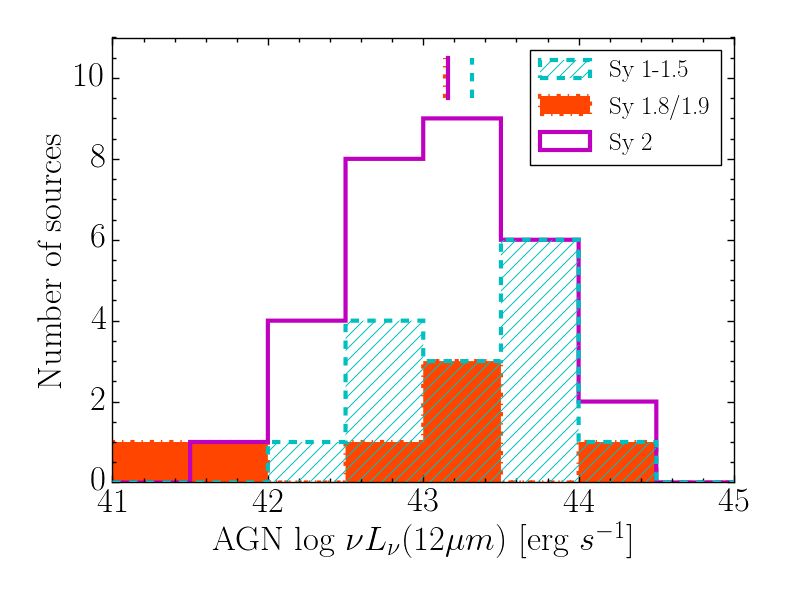}{\vspace{0cm}}
     
  
      \caption[AGN rest-frame 12\,$\mu$m luminosities distribution]
      {Derived AGN rest-frame 12\,$\mu$m luminosities distribution
      for 
      Seyfert 1-1.5 galaxies (hatched cyan histogram),
      Seyfert 1.8/1.9 galaxies (filled orange histogram) and
      Seyfert 2 galaxies (magenta histogram). 
      The vertical lines indicate the median of the distributions.} 
   
  \label{fig_12micras_luminosity}

  \end{center}
\end{figure}



\begin{figure*}
    \begin{center}

      \includegraphics[width=0.31\textwidth]{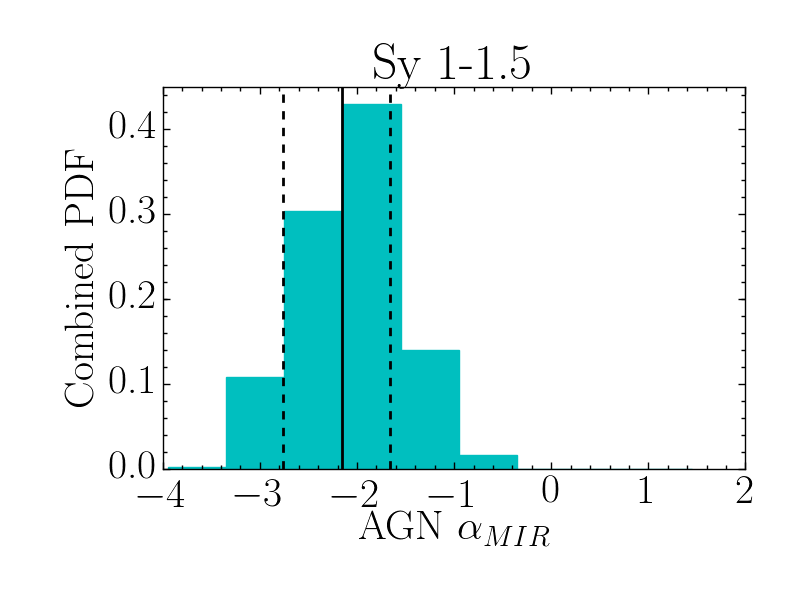}{\vspace{0cm}}
      \includegraphics[width=0.31\textwidth]{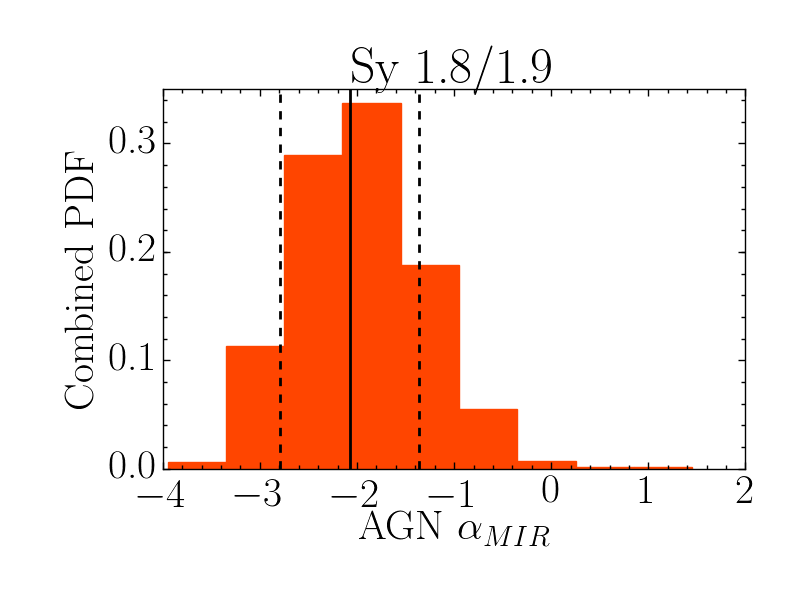}{\vspace{0cm}}
      \includegraphics[width=0.31\textwidth]{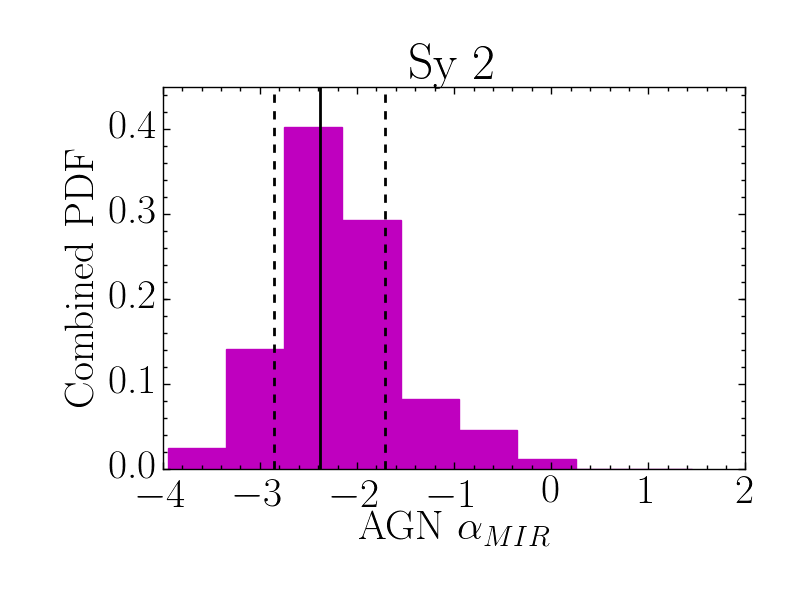}{\vspace{0cm}}

      \caption[Combined probability distribution functions of the AGN MIR spectral index]
      {Combined probability distribution functions of the AGN MIR ($8.1-12.5$\,$\mu$m)
  spectral index  derived
  with \textsc{deblend}IRS. In all panels the solid lines indicate the median of the distributions
  and the dashed lines the 16\% and 84\% percentiles. The panels are for the Seyfert 1-1.5 galaxies (cyan, left),
   Seyfert 1.8/1.9 galaxies (orange,  middle), and   Seyfert 2 galaxies (magenta, right).
  } 
   
  \label{figure_alpha_denblendIRS}

  \end{center}
\end{figure*}


\begin{figure*}
    \begin{center}

      \includegraphics[width=0.31\textwidth]{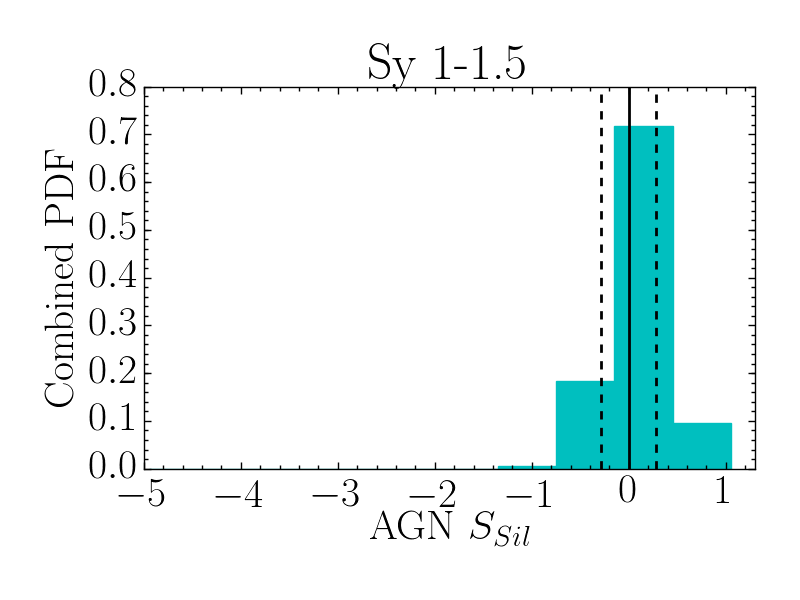}{\vspace{0cm}}
      \includegraphics[width=0.31\textwidth]{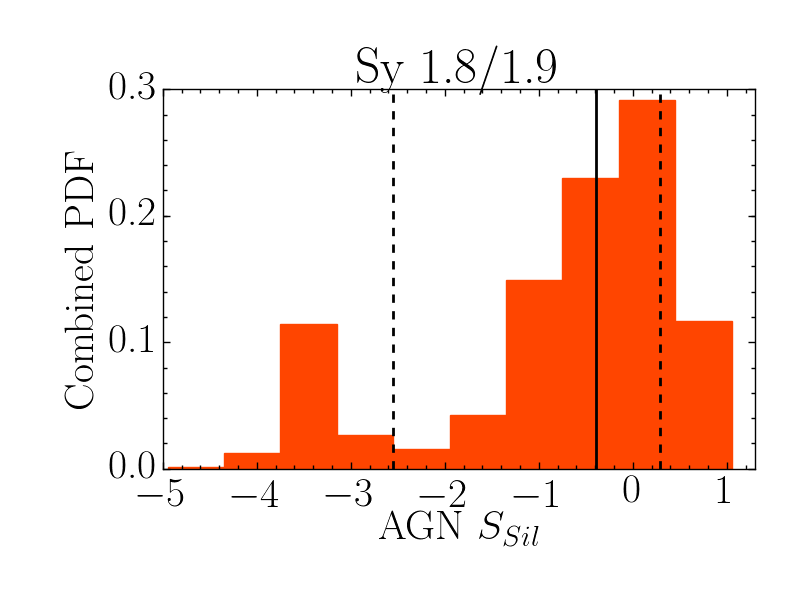}{\vspace{0cm}}
      \includegraphics[width=0.31\textwidth]{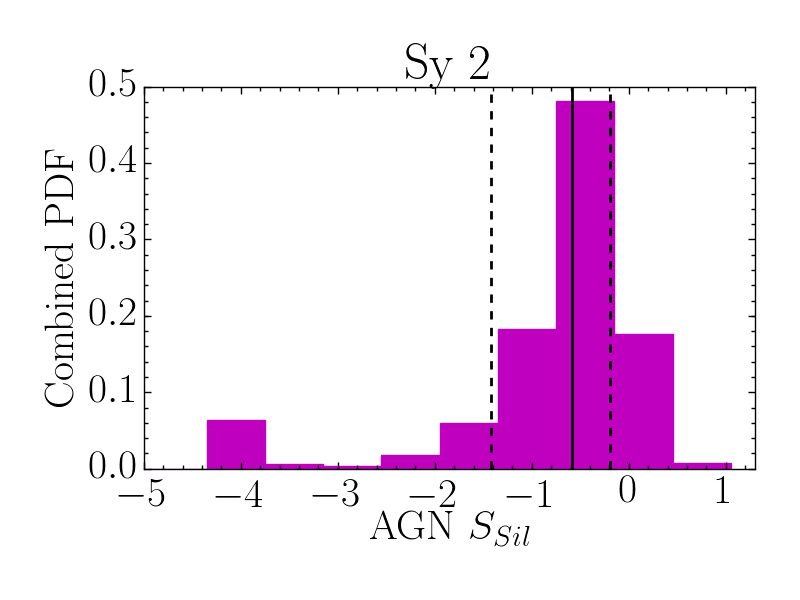}{\vspace{0cm}}

      \caption[Combined probability distribution functions of the AGN strength of the silicate feature]
      {Combined probability distribution functions of the  AGN strength of the silicate feature  derived
  with \textsc{deblend}IRS. In all panels the solid lines indicate the median of the distributions
  and the dashed lines the 16\% and 84\% percentiles.
  The panels are for the Seyfert 1-1.5 galaxies (cyan, left),
   Seyfert 1.8/1.9 galaxies (orange,  middle), and   Seyfert 2 galaxies (magenta, right).
  } 
   
  \label{figure_Ssil_denblendIRS}

  \end{center}
\end{figure*}


\begin{table*}
 \caption{Statistics of the combined probability
 distributions}
\label{table-statistics-varias-figuras}
\begin{center}
\resizebox{12cm}{!}{
\begin{tabular}{@{}lcccc}
\hline
Type   &  N &  AGN $\alpha_{\rm MIR}$  &  AGN $S_{\rm Sil}$  & AGN Fraction \\
\hline

Seyfert 1-1.5&15&-2.2 [-2.8, -1.7]&0.0 [-0.3, 0.3]& 0.82  [0.68, 0.96]\\
Seyfert 1.8/1.9 &7&-2.1 [-2.8, -1.4]&-0.4 [-2.6, 0.3]& 0.86  [0.75, 0.97]\\
Seyfert 2&30&-2.4 [-2.9, -1.7]&-0.6 [-1.4, -0.2]& 0.90  [0.76, 0.98]\\
Seyfert 2 (CAT3D models) & 19 & -2.6 [-2.9, -2.0]&-0.4 [-0.8, -0.1]& 0.88 [0.73, 0.96]\\

\hline

\end{tabular}}
\end{center}

Notes.-- The AGN MIR spectral index is estimated in the $8.1-12.5$\,$\mu$m range. 
The  AGN fractional contribution refers to the $5-15$\,$\mu$m luminosity.
We give the median value and in parenthesis the 16\% and 84\% 
percentiles.

 \end{table*}


\subsection{MIR properties of AGN}
\label{seccion-MIR-properties}
The first result from the spectral decomposition is that
  the AGN component dominates the MIR emission (over the $5-15\,\mu$m spectral range) 
on nuclear scales (typically 100-150\,pc), with a
        median value of 87\% for the full sample. 
        Moreover, the AGN contribution is similar for the Seyfert 2 and Seyfert 1-1.5 in
our sample  (median values of 88\% and 86\%, respectively).   The number of Seyfert 1.8/1.9 galaxies
is small for statistics, so the slightly difference in their median value (80\%) is not significant. 
We note,  however, that the physical sizes covered by the slits of the Seyfert
1-1.5 nuclei are larger than those of the Seyfert 2 nuclei, on
average. This means that if the slits were covering similar physical sizes for type 1-1.5 and 2 in our
sample, then the AGN fractional
contribution  in the MIR in Seyfert 1 nuclei should be slightly higher.
This is in agreement with the prediction of a nearly  isotropic emission at $12\,\mu$m
of the clumpy torus models of \cite{Nenkova2008b} and 
the similarity of the MIR emission of type 1 and
type 2 AGN when compared with their hard
  X-ray luminosity, which is a proxy for the AGN bolometric luminosity 
  \citep[see e.g.,][]{Alonso-Herrero2001,Krabbe2001,Lutz2004,Gandhi2009,Levenson2009,Asmus2015}.
  These observational differences in the MIR between type 1 and type 2
  are found only to be at most a factor of two \citep{Burtscher2015}.
      Accordingly, we also find from the spectral decomposition that the typical AGN
luminosities at rest-frame 12\,$\mu$m 
of Seyfert 1-1.5 galaxies (median $\log (\nu L_{12\mu{\rm m}}/{\rm
    erg\,s}^{-1})=43.3$) are only slightly higher than those of Seyfert 2
galaxies (median $\log (\nu L_{12\mu{\rm m}}/{\rm
  erg\,s}^{-1})=43.2$) in our sample (see also Fig.~\ref{fig_12micras_luminosity}).

In terms of the strength of the AGN silicate feature,  the Seyfert 2 and Seyfert 1.8/1.9
have a median value of
  $S_{\rm Sil}=-0.5$ and  $S_{\rm Sil}=-0.3$, respectively, whereas the Seyfert 1-1.5 typically show a flat or slightly
 in emission feature (median $S_{\rm Sil}=0.0$) and show a 
 narrower range of fitted values.  This indicates that the behaviour of the Seyfert 1.8/1.9 is closer to the
Seyfert 2 than to the Seyfert 1-1.5 in terms of the strength of the silicate feature.
 The difference in the strength of the silicate feature is a well known property
as Seyfert 1-1.5 generally show the silicate feature in emission and the Seyfert 2 galaxies
in absorption   (\citealt{Shi2006}, \citealt{Thompson2009}, \citealt{AAH2014}, but also see
\citealt{Hatziminaoglou2015}). 
However, the difference in the strength of the silicate feature is not necessarily reflecting
the properties of the torus because the MIR nuclear emission of some
Seyfert galaxies may be due to extended dust components in the host galaxy
\citep{Goulding2012} even on sub-arcsecond scales
\citep{Roche2006,Hoenig2010,AAH2011,
  GonzalezMartin2013,  AAH2014, AAH2016}. We will come back to this issue
when we compare the observations with the CAT3D torus model
predictions in Section~\ref{sec:CAT3Dvsobs}.
 The median values of the fitted AGN MIR spectral indices are
  $\alpha_{\rm MIR}=-2.1$, $\alpha_{\rm MIR}=-2.1$ and $\alpha_{\rm MIR}=-2.3$ for Seyfert 1-1.5, Seyfert 1.8/1.9 and Seyfert 2, respectively but the
  ranges are similar for all
  Seyfert types in our sample. 
The difference in the MIR spectral index  has also
  been noted by, among many works, \cite{Ramos-Almeida2011} who 
  found that Seyfert 2 show steeper $1-18\,\mu$m 
spectral energy distributions (SEDs) than Seyfert 1. However, they found that the difference in the $8-13\,\mu$m 
spectral range was small \citep[see also][]{AAH2014}.


%
%
%
%
%
%
%
%
%
%


\subsection{Comparison for different Seyfert types and other AGN}

To make a statistical comparison  of the MIR properties of Seyfert 2 and Seyfert 1-1.5 
we obtained the combined PDF of each of the subsamples as a simple average of the PDF of the individual
galaxies for the AGN MIR spectral index,
the strength of the silicate feature and the AGN fractional contribution within the slit
to the $5-15$\,$\mu$m luminosity.
Figure~\ref{figure_alpha_denblendIRS} shows 
that the peaks of the combined PDF of the AGN MIR spectral index  are similar for all the Seyfert types
although the distribution is slightly narrower for the Seyfert 1-1.5 nuclei
(see also the statistics in Table~\ref{table-statistics-varias-figuras}). 
However, the differences between different Seyfert types are more
apparent for the combined PDF of the silicate strength.
Not only the peaks of the combined PDFs are significantly different
for the three types (as also noted for the individual fits in the
previous section) but the distributions differ. The Seyfert 1.8/1.9 and 
Seyfert 2 nuclei peak at the feature in absorption and also show a broad tail towards
deep silicate absorptions whereas the Seyfert 1-1.5 show a 
narrow distribution peaking at $S_{\rm Sil} = 0.0$. This is clearly seen in Fig.~\ref{figure_Ssil_denblendIRS}.



We can compare the   AGN MIR properties of our Seyfert
  galaxies with those  of a IR-weak quasars
  (type 1) derived  by \cite{Alonso-Herrero2016b}. This sample includes 10 optically
    selected local quasars, mostly Palomar-Green (PG) quasars, with sub-arcsecond MIR spectroscopy 
    and for which the AGN MIR spectral properties were derived with a
    similar methodology as the Seyfert galaxies (Section~\ref{seccion-resultados-DeblendIRS}).
  These quasars are classified as IR-weak quasars based on
  their total IR to optical B-band luminosity ratios. We refer the reader to \cite{Alonso-Herrero2016b} for further details
  on the quasar sample.
  The combined PDFs of the IR-weak quasars \citep[see table~6 of][]{Alonso-Herrero2016b}
    have median values for the 
  AGN MIR spectral
  index of $\alpha_{\rm MIR}=-1.7$ ($1\sigma$ confidence interval of [$-2.4$, $-1.0$]) and for the
  strength of the silicate feature of $S_{\rm Sil}=0.1$ ($1\sigma$ confidence interval of [$-0.2$, 0.3]). 
Therefore,  the  quasars have significantly flatter AGN MIR spectral indices
  than the Seyfert 1-1.5 ($\alpha_{\rm MIR}=-2.1$) but similar strengths of the silicate feature.
 In Section~\ref{sec:CAT3Dvsobs} we will
use clumpy torus model predictions to see whether
  these differences also imply differences in the torus properties for the different types of AGN.

%
%

\section{Statistical comparison with the CAT3D clumpy torus
models}
\label{models}

\subsection{Brief description of the CAT3D models}
\label{sec-differences-new-old-models}

\begin{figure}
    \begin{center}
      
      \includegraphics[width=0.49\textwidth]{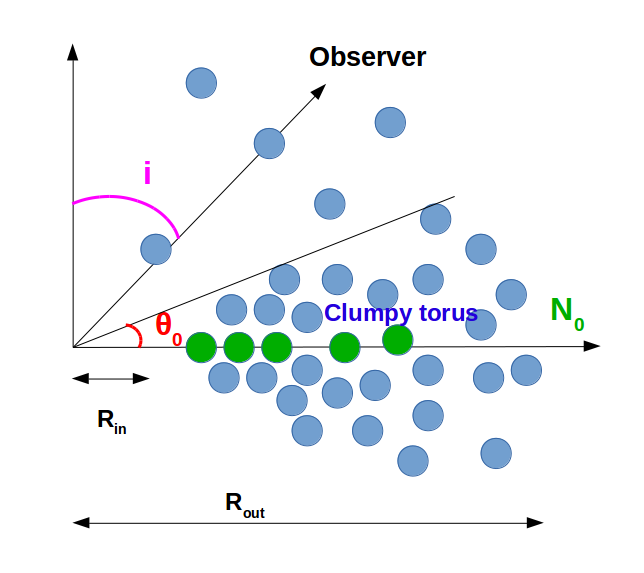}{\vspace{0cm}}
  
  \caption[Representation of the CAT3D clumpy torus parameters]{Representation of the CAT3D clumpy torus showing some of the parameters that
  characterise the models, namely the half-covering angle $\theta_0$, in red; the inclination $i$, in magenta,
  the number of clouds  along an equatorial line-of-sight $N_0$, in green; and the inner and outer torus
  radii $R_{\rm in}$ and $R_{\rm out}$ respectively, in black.
  }

  \label{fig-parametros-toro}

  \end{center}
\end{figure}

In this work we make use of the \cite{Hoenig-kishimoto} CAT3D clumpy torus 
models that provide the model SED for clumpy dust emission in a torus
around the AGN accretion disk. These models are characterised by six
parameters that have direct influence on the  IR dust SEDs of AGN. These are:  (1)
the  power-law index of the radial dust-cloud distribution $a$, that is $\propto r^a$; (2) the half-covering angle
of the torus $\theta_0$; (3) the number
of clouds along an equatorial line-of-sight $N_0$; (4) the torus outer radius $R_{\rm out}$; (5) the optical depth of
the individual clouds $\tau_{\rm V}$; and (6) the inclination ( i.e., the viewing angle) $i$. In
Fig.~\ref{fig-parametros-toro} we show a sketch
of some of the CAT3D torus model parameters. The AGN is assumed to be radiating in an isotropic manner. However,
in Section~\ref{sec:anisotropicAGN} we will investigate very briefly the effects of introducing anisotropic AGN
  radiation on the predicted MIR properties of the clumpy torus models. 


\begin{table*}
 \caption{Parameters of the CAT3D clumpy torus models.}
 \label{tabla-parametros-modelos}
\begin{center}

\resizebox{14cm}{!}{
\begin{tabular}{@{}lccc}
\hline
  
Parameter & Symbol  &Old models&New models\\
 \hline
Index cloud radial distribution & $a$& [0.00, -2.00] steps of 0.5& [0.50, -1.75] steps of 0.25\\
Torus half-covering angle & $\theta_0$&30\textdegree, 45\textdegree, 60\textdegree, 85\textdegree&30\textdegree, 45\textdegree, 60\textdegree\\
Clouds along equatorial direction & $N_0$&2.5, 5.0, 7.5, 10.0&2.5, 5.0, 7.5, (10.0, 12.5)$^*$\\
Cloud optical depth &  $\tau_{\rm V}$&30, 50, 80&50\\
Torus outer radius & $R_{\rm out}$&150&450\\
Inclination & $i$& [0\textdegree, 90\textdegree] steps of 15\textdegree & [0\textdegree, 90\textdegree] steps of 15\textdegree\\ 
\hline

\end{tabular}}

Notes--- The outer radius is measured in units of the sublimation radius.
$^*$The values of $N_0=10, \,12.5$ are only for models with
$a\le 0$ for computational reasons. 
\end{center}
\end{table*}

To calculate the IR SEDs of these models several steps are carried out.
The first step is to simulate each cloud by Monte Carlo radiative transfer simulations.
Then, dust clouds are randomly distributed around the AGN, according to the physical
and geometrical parameters, each one associated with a model cloud from the first step.
The final torus SED is calculated via raytracing
along the line-of-sight from each cloud to the observer.
This method allows to take into account the three dimensional
distribution of clouds and the statistical variations of randomly distributed clouds.
 We refer the reader to \cite{Hoenig-kishimoto} for a complete description of the calculations.


\subsection{New CAT3D runs with a physical dust sublimation model}
\label{dustsublimationmodel}
The CAT3D clumpy torus models published by \cite{Hoenig-kishimoto} (hereafter old models)
assumed a standard ISM composition for the dust, containing 47\% of graphite
and 53\% of silicates. In this work we present a new version of the models (hereafter
referred to as new models) that includes additional more realistic physics in an
attempt to model the differential dust grain sublimation. Graphite grains can sustain higher 
temperatures than silicate grains, with the former being able to heat up
to $\sim$ $1900-2000$\,K and the latter sublimating at $\sim$ $800-1200$\,K, depending on density  
\citep{Phinney1989}.

The new sublimation model assumes that silicates are sublimated away once their temperature goes above 1250\,K. 
This means that all those clouds at distances from the AGN as to heat up to temperatures above 1250\,K will not
contain any silicates.  Therefore, in the new models the absorption and scattering efficiencies are adjusted
  accordingly. 
In addition, the hottest dust at $T\simeq 1900\,$K will only contain larger graphite grains, 
that is, the minimum grain size for the ISM dust size distribution is increased from 0.025\,$\mu$m 
to 0.075\,$\mu$m. This accounts for the fact that small grains are cooling less efficiently 
and will reach the sublimation temperature at larger distances than larger grains.
 As we shall see, since graphites have higher emissivity, this will result in bluer 
 NIR to MIR SEDs for a
given set of torus model parameters
than in the old models, which had a standard ISM dust  composition without a sublimation model.
We note that  the different sublimation temperatures included in the new models are unique
to these models and are not taken into account in other available clumpy 
 torus models \citep[e.g.,][]{Schartmann2008,Nenkova2008a,Nenkova2008b},
although  smooth torus models have used various approaches to 
differential sublimation \citep[e.g.,][]{GranatoDanese1994,Schartmann2005,Fritz}.


\begin{figure*}
 \centering   
      \includegraphics[width=1\textwidth]{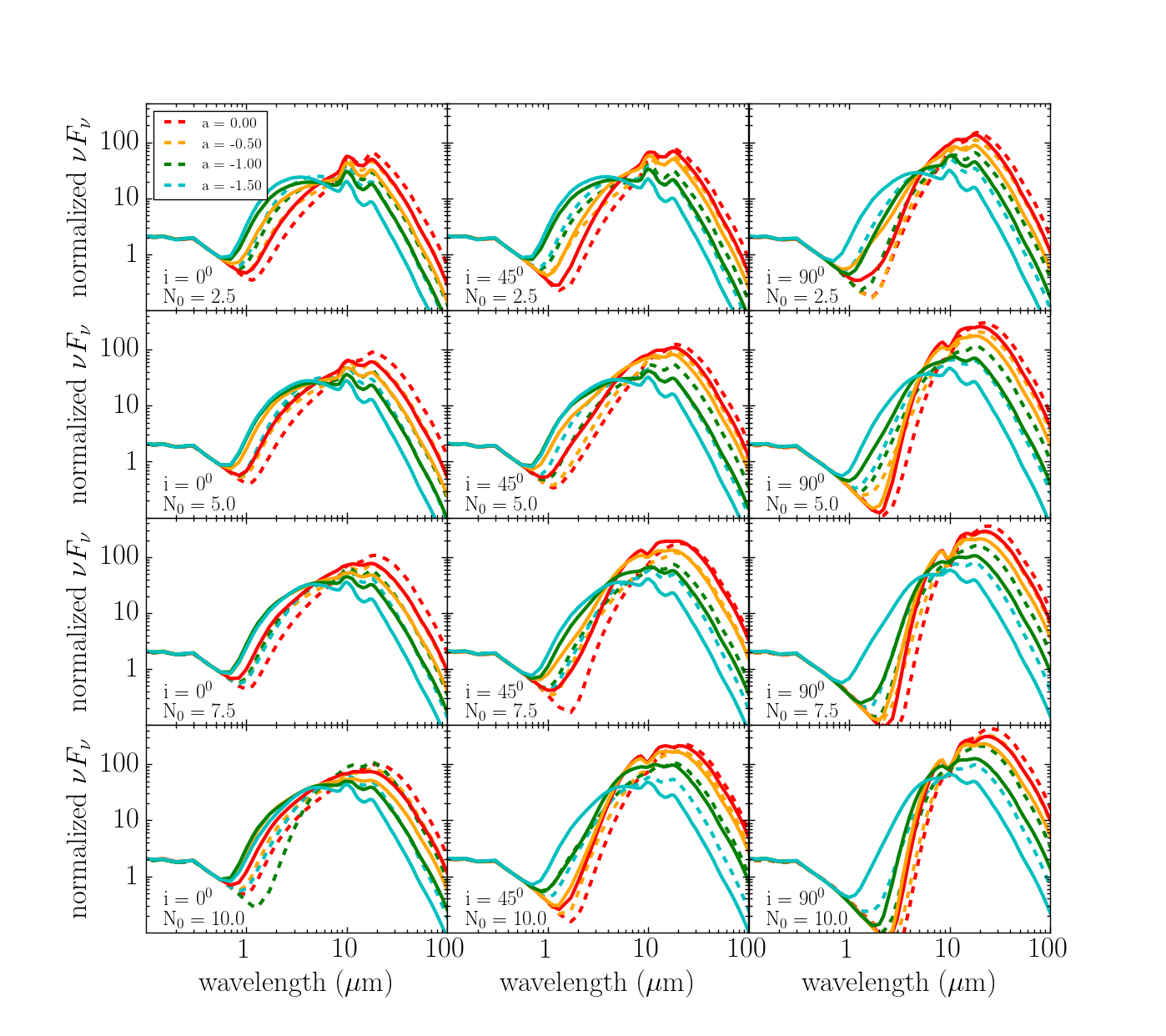}{\vspace{0cm}}
       \vspace{-1cm}    
  
  \caption{Examples of the CAT3D clumpy torus SEDs for the old models (dashed lines)
  and new models  (solid lines) normalised at 0.5\,$\mu$m. 
  From top to bottom, the rows show an increasing number 
  of clouds along an equatorial line-of-sight,  $N_0= $2.5, 5.0, 7.5, and 10.0, respectively.
  The left column shows models for an inclination of
  0\textdegree, the middle column is for $i=45$\textdegree~and the right column is for $i=90$\textdegree. 
  In each panel we show the SEDs for one random cloud distribution for $a = 0.00$ (red), $a=-0.50$ (orange),
  $a=-1.00$ (green), and $a=-1.50$ (cyan).
  They all have  fixed values of  $\tau_{\rm V} = 50$ and $\theta_0 = 45$\textdegree.}

  \label{fig-SEDS-modelos-antiguos+nuevos}

\end{figure*}



\begin{figure*}
    \centering    
      \includegraphics[width=1\textwidth]{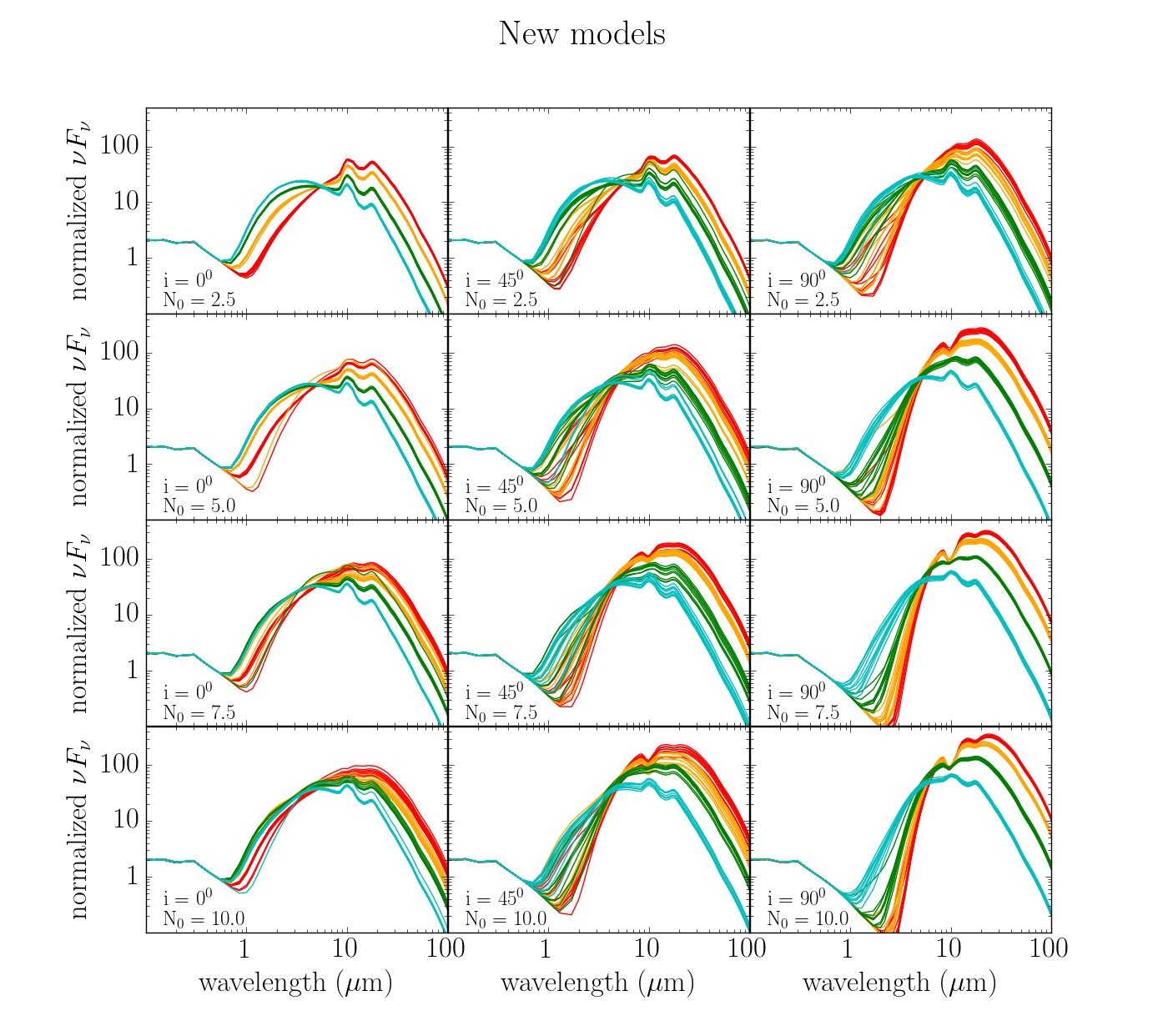}{\vspace{0cm}}
       \vspace{-1cm}    
  
       \caption{Same as Fig.~\ref{fig-SEDS-modelos-antiguos+nuevos} but  only for the new models and
         showing ten random distributions
         of clouds for 
  each configuration of parameters.}

  \label{fig-SEDS-modelos-nuevos-todas-distribuciones}

\end{figure*}


The old and the new models cover different ranges of the torus parameters. In Table~\ref{tabla-parametros-modelos} 
we summarize the parameter values and ranges used in the old and new models. The index of the radial
distribution of clouds $a$ covers different ranges, namely 
[0.00, -2.00] for the old models and [0.50, -1.75] for the new ones, and different steps of 0.50 and 0.25, respectively.
We note that for the new models we are adding inverted radial cloud distributions (positive values of $a$).
Although, probably not very common, inverted radial distributions resemble a disk-like accretion flow that thins out
towards the inner radius. This may be the kind of geometry needed to explain the population of hot dust poor quasars \citep{Hao2010}.
In the case of  the half-covering angle of the torus $\theta_0$, the old models provided more values 
($\theta_0=$30\textdegree, 45\textdegree, 60\textdegree, 85\textdegree)
than the new ones ($\theta_0=$30\textdegree, 45\textdegree, 60\textdegree). For  the number of clouds 
$N_0$ the ranges are also different, [2.5, 10.0] for the old models and
[2.5, 12.5] for the new models, both in steps of 2.5.

 The new and old CAT3D models have different values of the torus outer radius $R_{\rm out}$ 
(measured in units of the sublimation radius, see Table~\ref{tabla-parametros-modelos})
due to the smaller dust sublimation radius of the graphite/large 
dust grains than the silicate grains. The grain sublimation radius is proportional to the square root
  of the AGN luminosity. For $L_{\rm AGN}=10^{46}\,{\rm erg\,s}^{-1}$, the sublimation radius is 0.5\,pc for large
  grains and 0.955\,pc for the typical ISM dust composition 
\citep[see table~2 of][]{Hoenig-kishimoto}. Also, the
  value of the outer radius of the torus needs to be sufficiently large so that it encompasses the physical sizes of the torus
  measured at all wavelengths and was chosen to have the same size range as with the old models. 
In the old models we have three values of   cloud optical depth $\tau_{\rm V}$, 30, 50, and 80,
whereas the new models only have  one value  $\tau_{\rm V}$=50.
The range of  the inclination $i$ is the same for both the old and the new models,
from 0\textdegree~to 90\textdegree~in steps of 15\textdegree.

As explained above, the old and new models cover different ranges of torus
parameters. We have a total of 1680  parameter configurations for old models and 966 configurations for new models, 
 with 336 configurations sharing the same values of the torus model parameters.
For each configuration of parameters, we have one SED obtained from a random arrangement
of clouds for the old models and ten 
for the new ones, obtained from ten random distributions of the clouds satisfying 
the same configuration of parameters.

In Fig.~\ref{fig-SEDS-modelos-antiguos+nuevos} we show some  examples of SEDs (in units of $\nu f_\nu$) for the 
old and the new models in dashed and solid lines respectively, for a random cloud distribution. Each column represents
a different inclination ($i = 0$\textdegree, 45\textdegree, and 90\textdegree) and each row different values of the number of clouds 
in the equatorial line-of-sight ($N_0$ = 2.5, 5.0, 7.5, and 10.0). They all have 
fixed values of  $\tau_{\rm V} = 50$ and $\theta_0 = 45$\textdegree. Each panel shows four different values
of the power-law index of the radial dust-cloud distribution, $a$, indicated with different colours. 
For both the old and new models the total SEDs becomes redder for flatter 
 radial cloud distributions (more positive
values of $a$). As explained in \cite{Hoenig-kishimoto}, this is because  flat power-law radial distributions have more
cool dust at larger radial distances.
There are differences in the continuum shape depending of the value of $a$.
For the flat and nearly flat distributions ($a = 0.0$ and $a=-0.50$), the continuum peaks at  longer
wavelengths than for the steeper distributions ($a = -1.0$ and $a=-1.50$). This is because 
in the steeper distributions there is more dust at small radial distances from the AGN so the 
average dust temperature is higher \citep{Hoenig-kishimoto}. This trend  does not depend 
on the inclination.

The $a$ values also have an effect in the strength of the 
silicate feature. For the torus parameters represented in
Fig.~\ref{fig-SEDS-modelos-antiguos+nuevos}  with the steepest
  radial distribution of clouds ($a = -1.5$), the silicate feature is always in emission, 
whereas for the rest of the $a$ values the strength of the feature
depends on the values of $N_0$ and the inclination.
While the SEDs have a substantial dependence on $a$, the dependence on $N_0$ is small. This 
dependence on $N_0$ is more important for the strength of the silicate feature
than for the shape of the SED (which is also related to $\alpha_{\rm MIR}$, see next section).  The 
silicate feature in emission is the strongest for $N_0$ = 2.5, whereas the feature
becomes less prominent  when there are more clouds along the 
equatorial direction (larger values of $N_0$). In the cases of the silicate in absorption, more clouds $N_0$
and higher inclinations result in  deeper absorptions.
 This is because changing $N_0$ has the effect that the inner and hotter part of the dust 
distribution becomes more obscured and more blocked along any light-of-sight.
 As a result, the effectively visible clouds are further away from the AGN and, thus, cooler. 
 These cooler clouds are more likely to show silicate absorption in their source functions.
 This effect, which is a particularity of a clumpy distribution, is combined with the fact that 
 the self-obscuration/extinction within the dust distribution reduces silicate emission features
 or turns them into absorption features.
As expected, the new models have bluer NIR to MIR SEDs for a
given set of torus parameters than the old
models. The differences in the shape are more noticeable for the steepest radial cloud distributions
($a = -1.5$), as there is more dust at small distances from the AGN, containing
only graphite for the new models while the old models
have silicates and 
graphites at the same distance. The differences also 
increase when there are more clouds along the 
equatorial direction (larger values of $N_0$) and for
more inclined views.

 In Fig.~\ref{fig-SEDS-modelos-nuevos-todas-distribuciones} we show the SEDs
for the new models, as in Fig.~\ref{fig-SEDS-modelos-antiguos+nuevos}, but representing
the ten random realisations of clouds computed for  each configuration of torus model parameters. The differences
between the random distributions for the same parameters are more apparent for more inclined views
and for the flattest radial distributions of the torus clouds ($a = 0.0$ and $a=-0.50$). This figure shows the 
importance of doing several random realisations instead of using only one.
 This is further discussed in Section~\ref{Section:CAT3D_vs_MIR_emission}.
 
\subsection{CAT3D predictions for the MIR emission}
\label{Section:CAT3D_vs_MIR_emission}

In this section we present the CAT3D torus model predictions for the
MIR emission of AGN and in particular for the properties
we analysed in Section~\ref{seccion-MIR-properties}, 
namely the MIR spectral index and the
strength of the silicate feature.
As explained by \cite{Hoenig-kishimoto}, although the angular size of the torus,  $\theta_0$, 
could be an additional source of degeneracy, there is a strong
relation between the index of the dust radial distribution, $a$ and the MIR spectral index.
There is also a strong relation between the number of clouds along the
equatorial direction, $N_0$, and the strength of the silicate
feature, even though the strength of the silicate feature also depends
on $a$ and  $\tau_{\rm V}$.
Using the clumpy torus models of  \cite{Nenkova2008a,Nenkova2008b}, 
\cite{Ramos-Almeida2014b} investigated the sensitivity of different observations in the
near and MIR to these torus model parameters. Specifically, they
found that  a detailed modelling of the $8-13$\,$\mu$m spectroscopy
  (not only the spectral index and strength of the silicate feature) of Seyfert galaxies can constrain
reliably the number of clouds and their optical depth.

We measured for each model SED the MIR spectral index
using 8.1 and 12.5\,$\mu$m as anchor points and 8 and 14\,$\mu$m to
fit the continuum and 10\,$\mu$m for the peak of the silicates, as the CAT3D models have both, the 
emission and the absorption features centred at 10\,$\mu$m \citep{Hoenig2010}.
For the old models,  for each of  the 1680 configurations  we obtained one value of the spectral index
and the strength of the silicate feature. For the new models we  measured ten values for
each  of the 966 configurations, which allows us to estimate the average and the standard deviation for the ten
values of the spectral index and the strength of the silicate feature for each configuration of parameters.
The typical scatters in the measured $\alpha_{\rm MIR}$ are $0.02-0.06$, although in the case of
a radial distribution index $a=0.5$ the scatter can be as high as 0.2. The typical scatter in the measured
$S_{\rm Sil}$ is $0.01-0.04$.
 These scatters are lower than the $1 \sigma$ uncertainties (16\% and 84\% percentiles) obtained with \textsc{deblend}IRS 
for the AGN MIR spectral index and the 
strength of the silicate feature distributions for the Seyfert galaxy sample (see Table~\ref{table-deblendIRS_results}).


\begin{figure*}

      \includegraphics[width=0.31\textwidth]{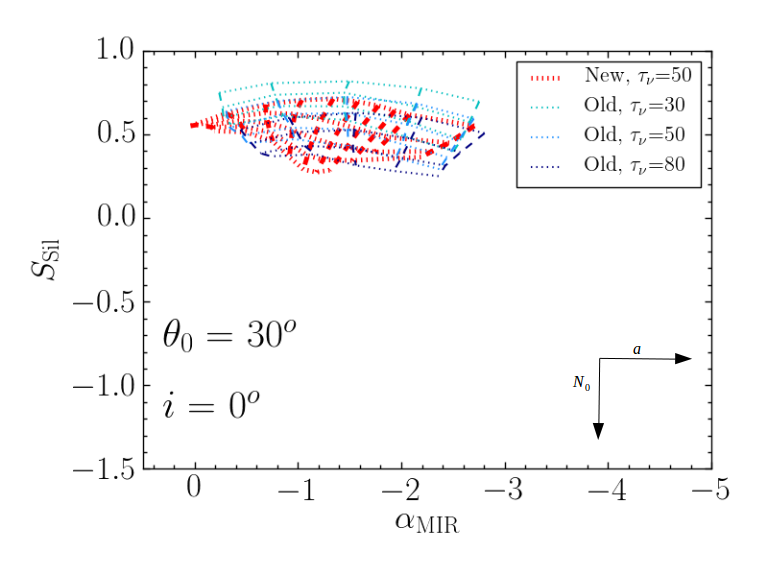}{\vspace{0cm}}
      \includegraphics[width=0.31\textwidth]{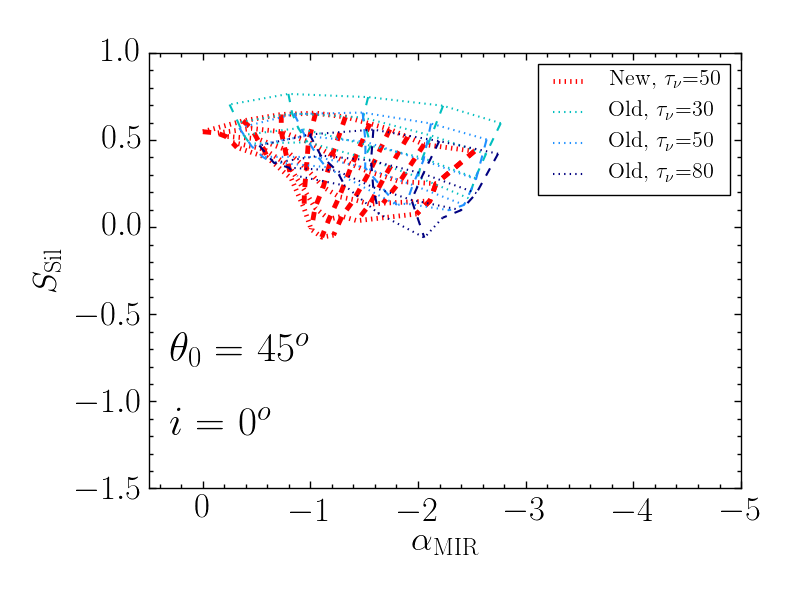}{\vspace{0cm}}
      \includegraphics[width=0.31\textwidth]{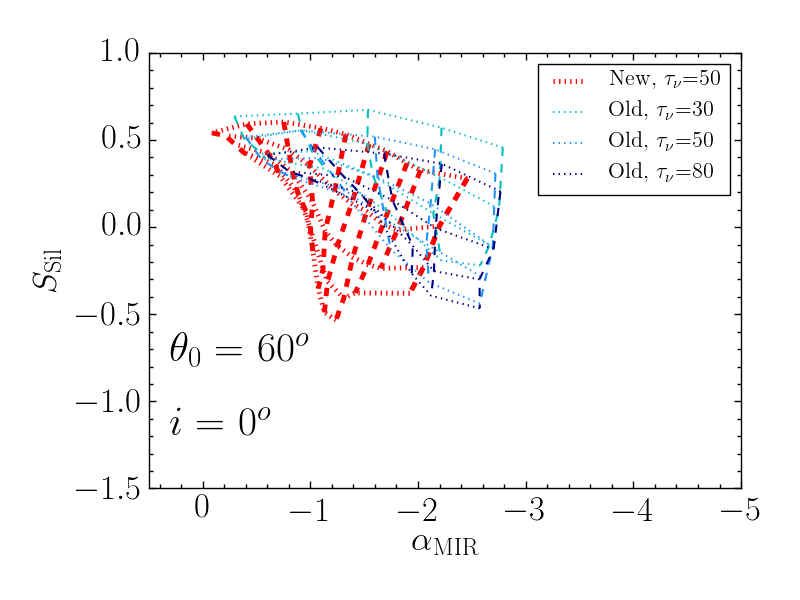}{\vspace{0cm}}      
      \includegraphics[width=0.31\textwidth]{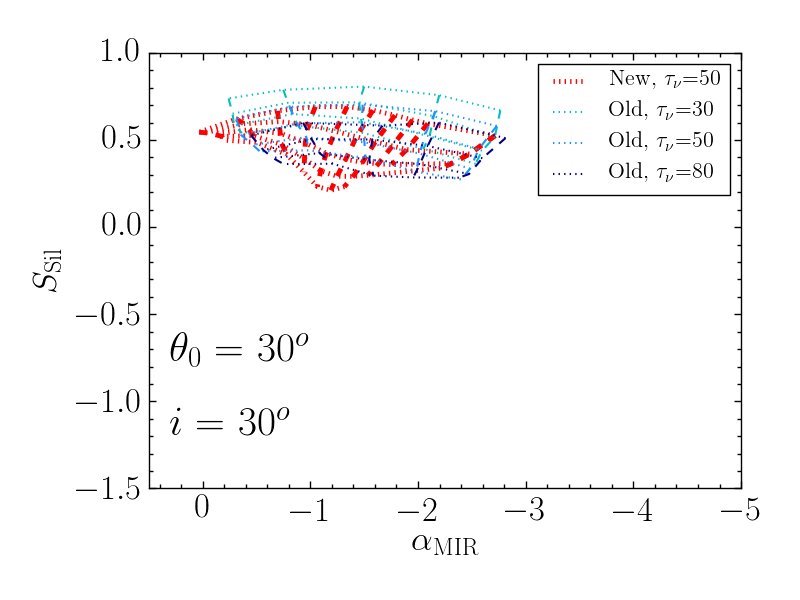}{\vspace{0cm}}
      \includegraphics[width=0.31\textwidth]{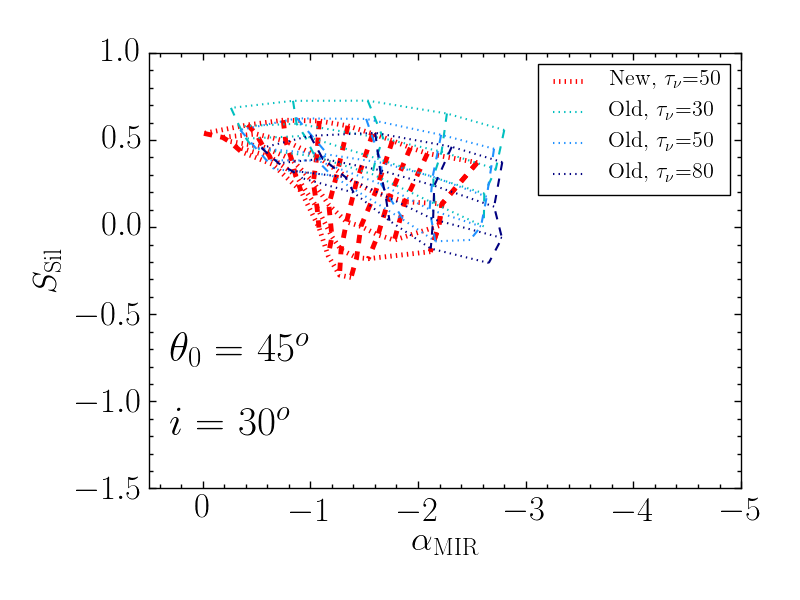}{\vspace{0cm}}
      \includegraphics[width=0.31\textwidth]{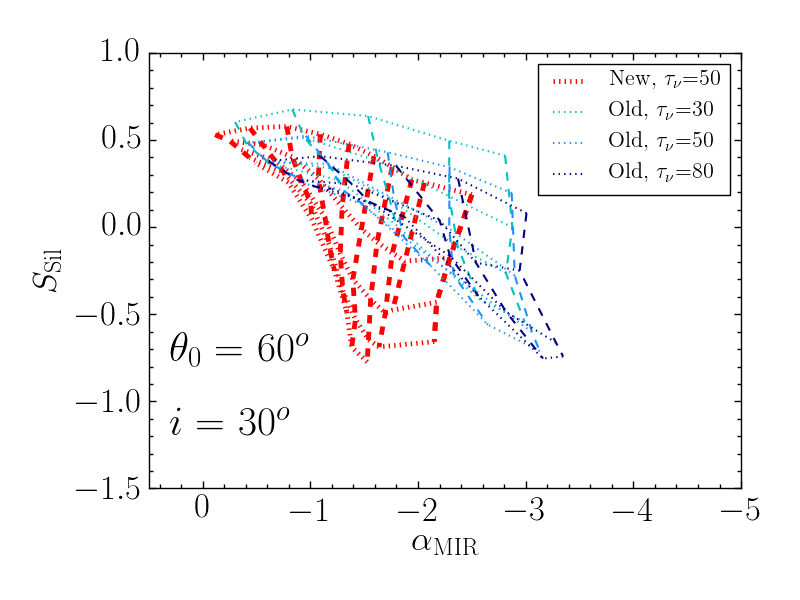}{\vspace{0cm}}   
      \includegraphics[width=0.31\textwidth]{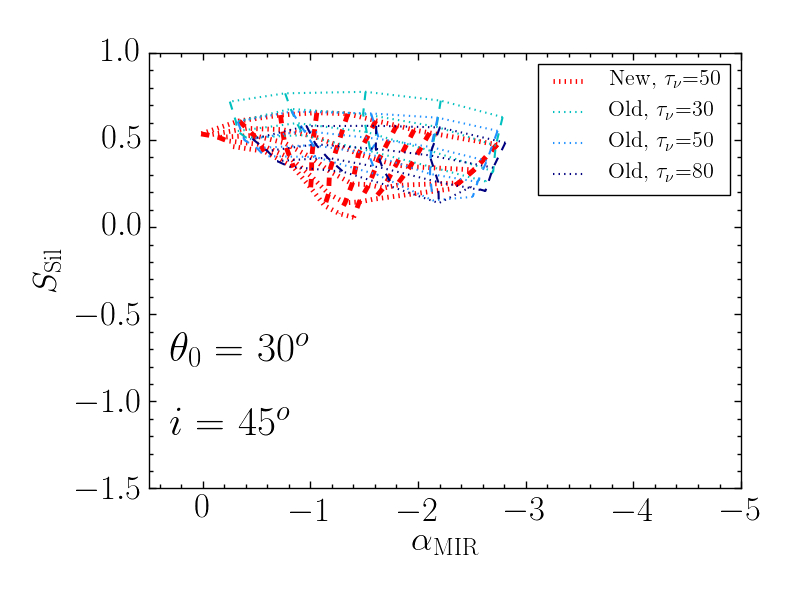}{\vspace{0cm}}
      \includegraphics[width=0.31\textwidth]{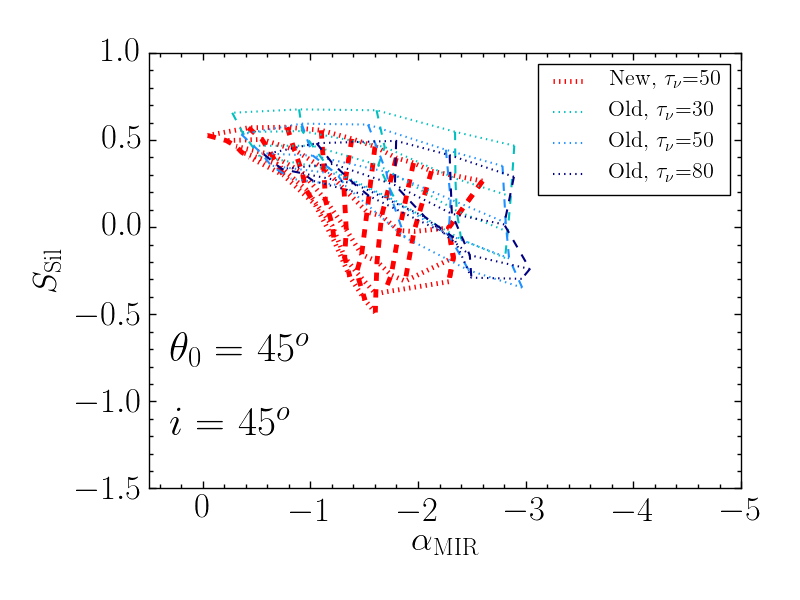}{\vspace{0cm}}
      \includegraphics[width=0.31\textwidth]{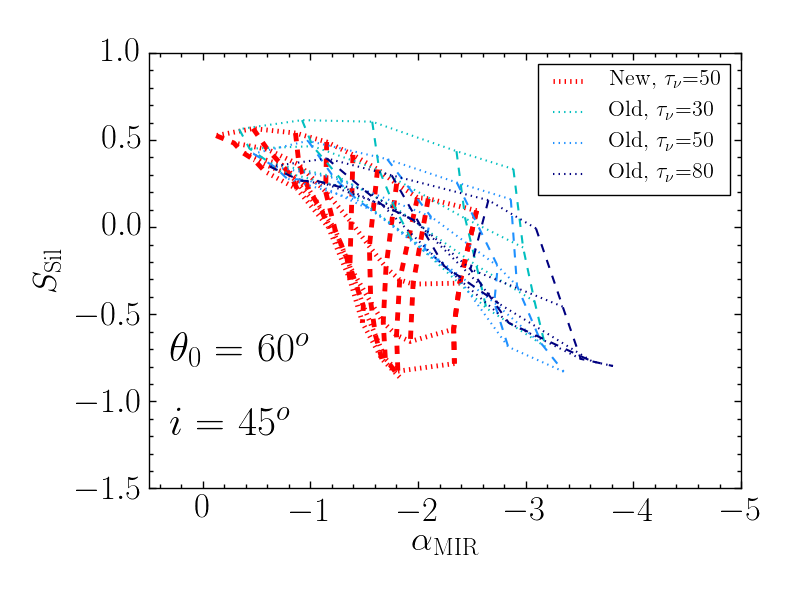}{\vspace{0cm}}    
      \includegraphics[width=0.31\textwidth]{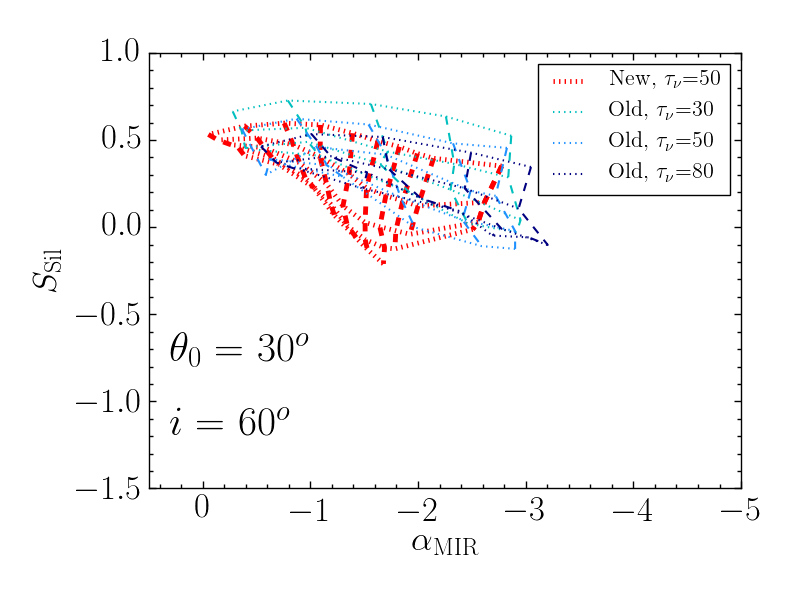}{\vspace{0cm}}
      \includegraphics[width=0.31\textwidth]{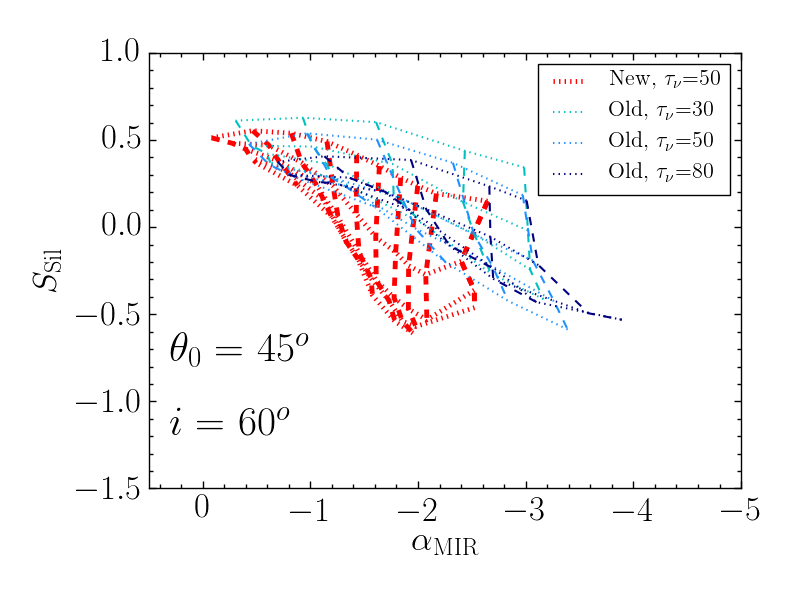}{\vspace{0cm}}
      \includegraphics[width=0.31\textwidth]{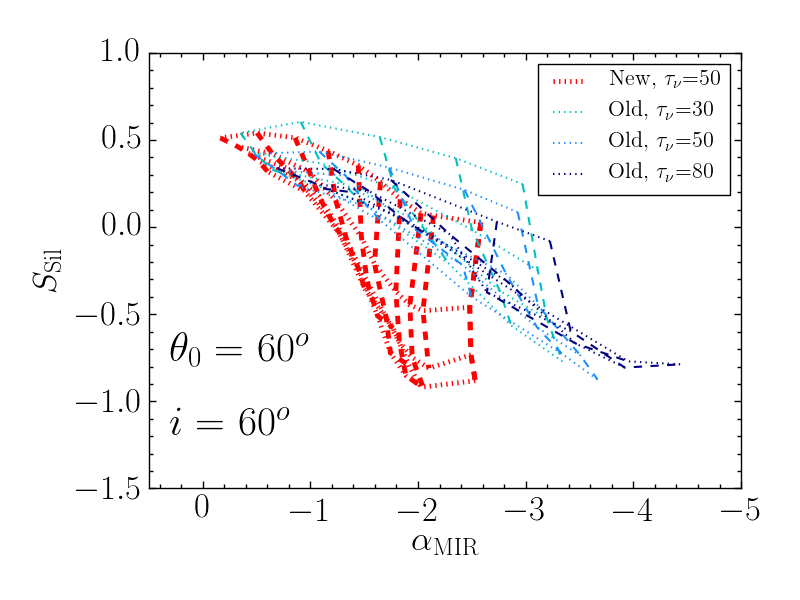}{\vspace{0cm}}   
      \includegraphics[width=0.31\textwidth]{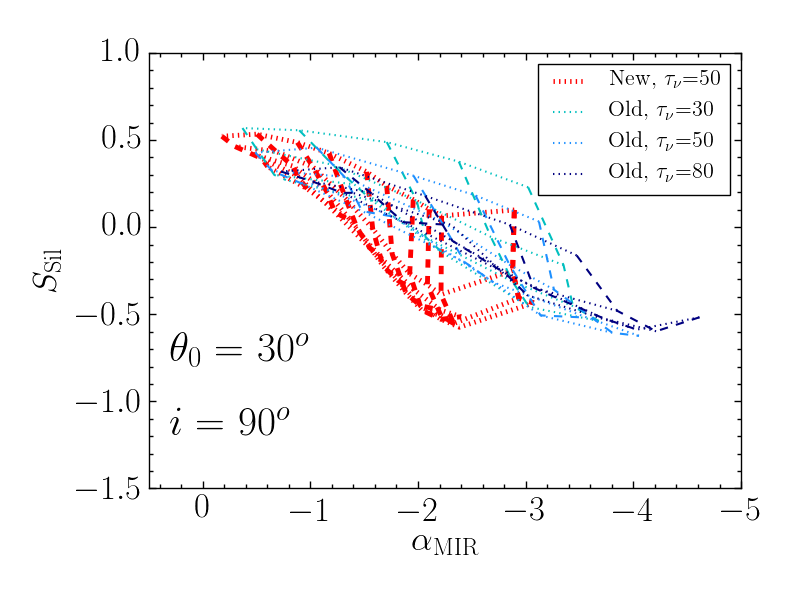}{\vspace{0cm}}
      \includegraphics[width=0.31\textwidth]{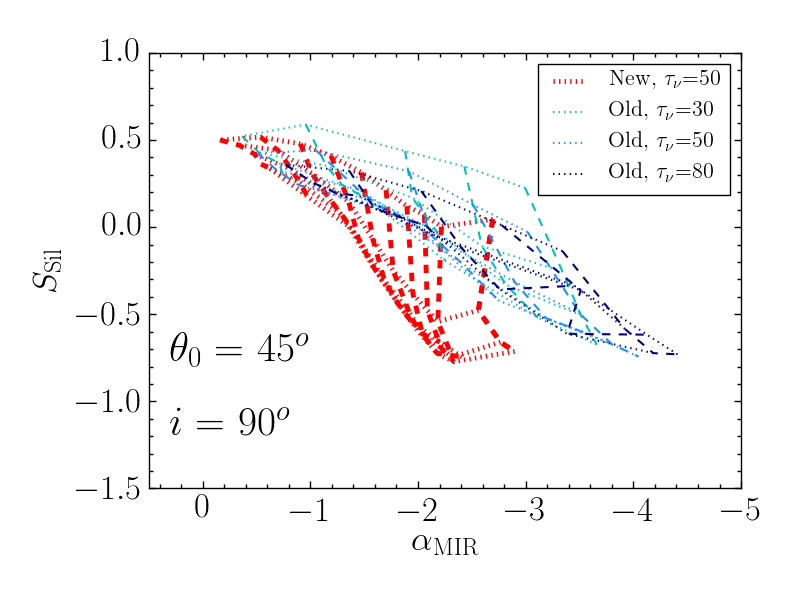}{\vspace{0cm}}
      \includegraphics[width=0.31\textwidth]{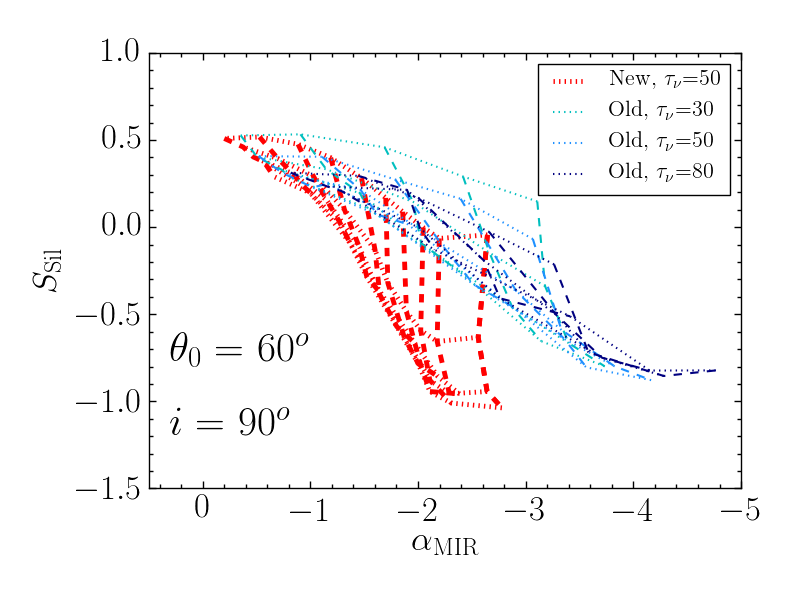}{\vspace{0cm}}

    \caption[Strength of the silicate feature against
      the MIR spectral index for the CAT3D models]{Strength of the silicate feature against
      the MIR spectral index for the CAT3D old models (in light blue  $\tau_{\rm V}=30$, medium
      blue  $\tau_{\rm V}=50$, and dark blue  $\tau_{\rm V}=80$),
    and the CAT3D new models (in red,  $\tau_{\rm V}=50$). Each panel shows the estimated values the full range 
    of $a$ (dashed lines,  values of $a$ becoming more negative to the left) and $N_0$ (dashed lines, 
    larger values moving down) for a fixed inclination and torus half-covering angle.
    Results are shown for five viewing angles $i=0$\textdegree, 30\textdegree, 45\textdegree, 60\textdegree, and 90\textdegree.} 
   \label{figura-comparacion-modelos}

\end{figure*}
   


\begin{figure*}

      \includegraphics[width=0.31\textwidth]{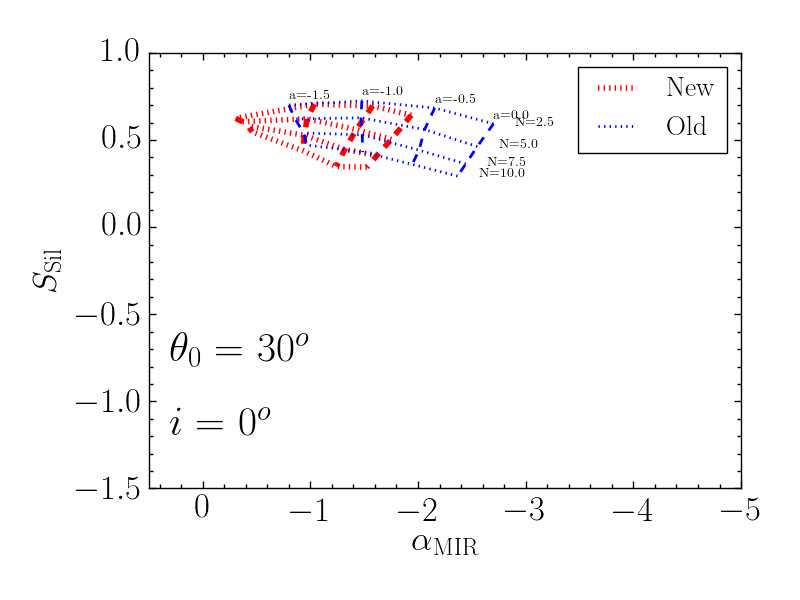}{\vspace{0cm}}
      \includegraphics[width=0.31\textwidth]{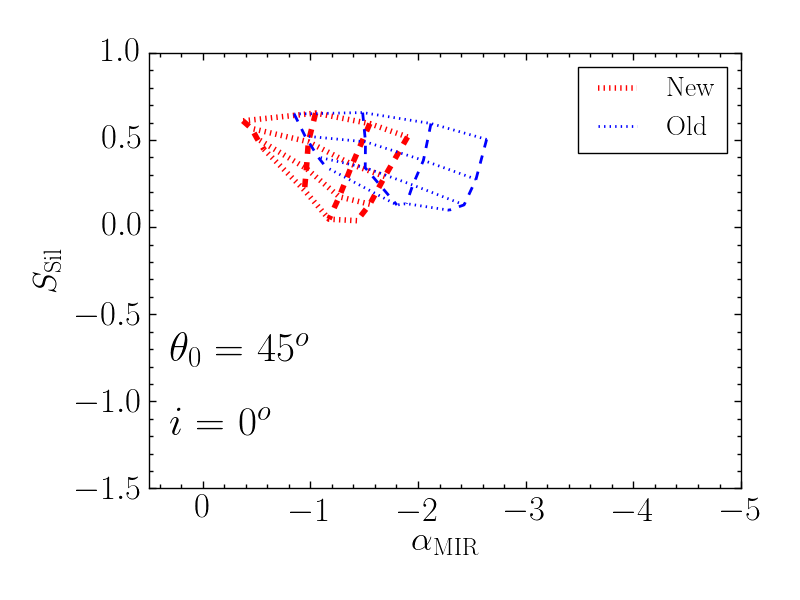}{\vspace{0cm}}
      \includegraphics[width=0.31\textwidth]{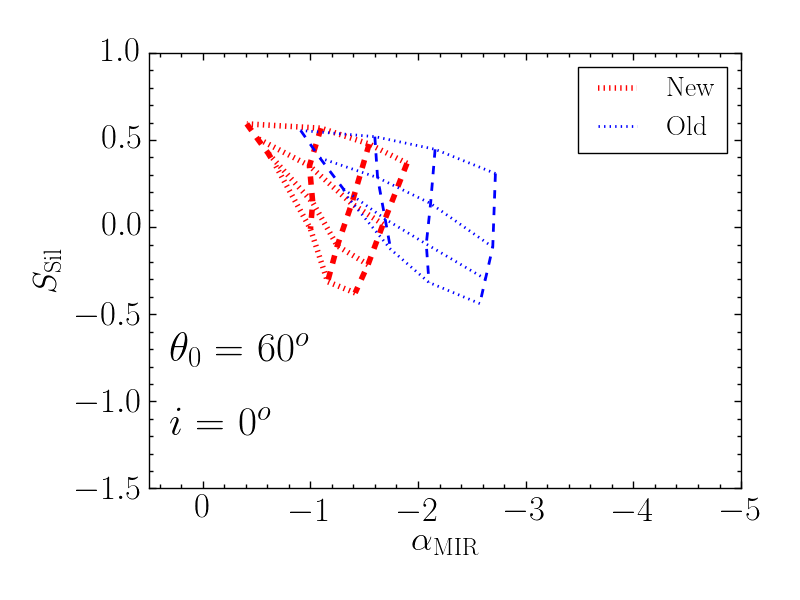}{\vspace{0cm}}       
      \includegraphics[width=0.31\textwidth]{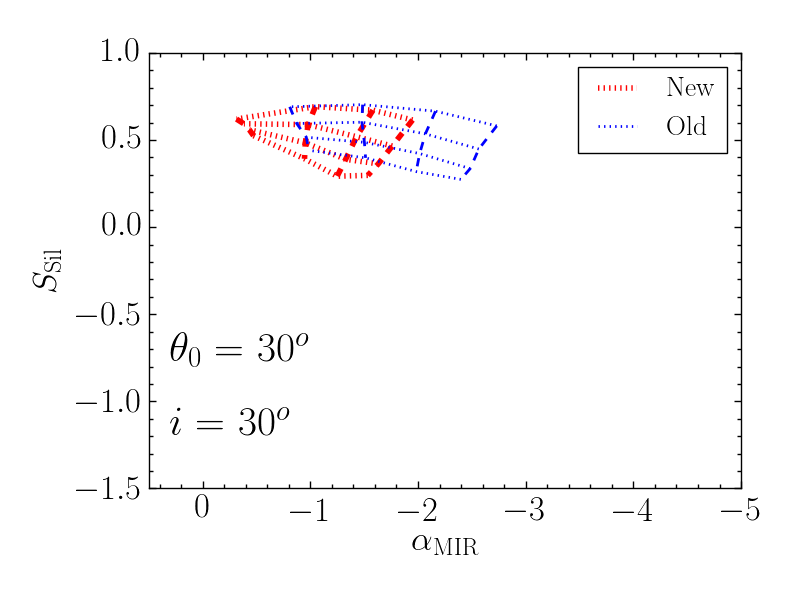}{\vspace{0cm}}
      \includegraphics[width=0.31\textwidth]{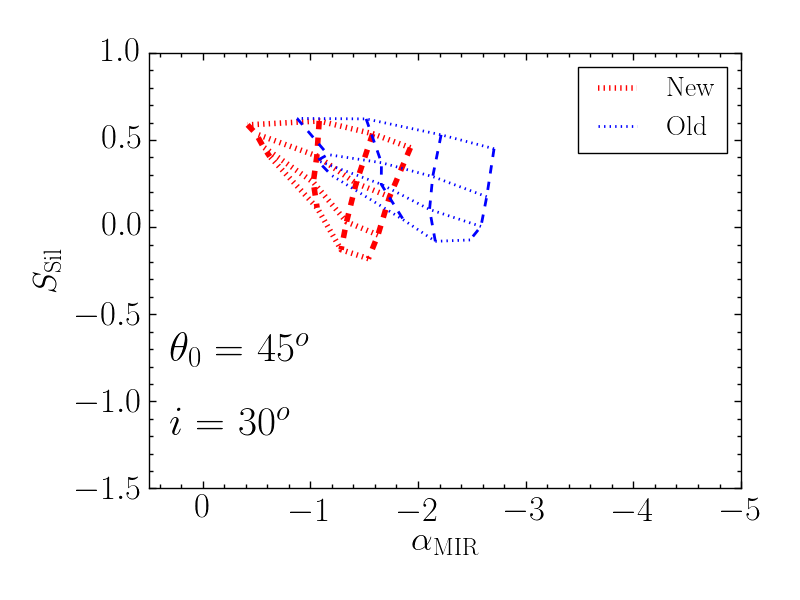}{\vspace{0cm}}
       \includegraphics[width=0.31\textwidth]{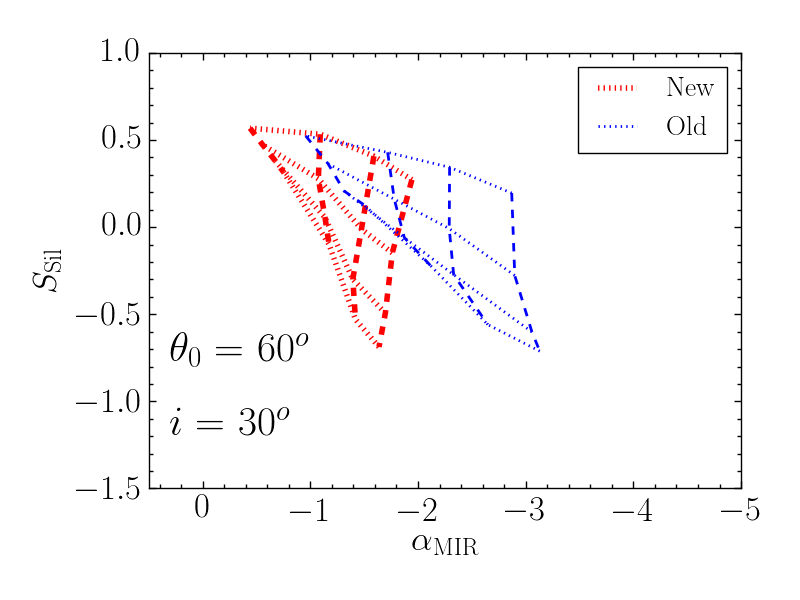}{\vspace{0cm}}    
      \includegraphics[width=0.31\textwidth]{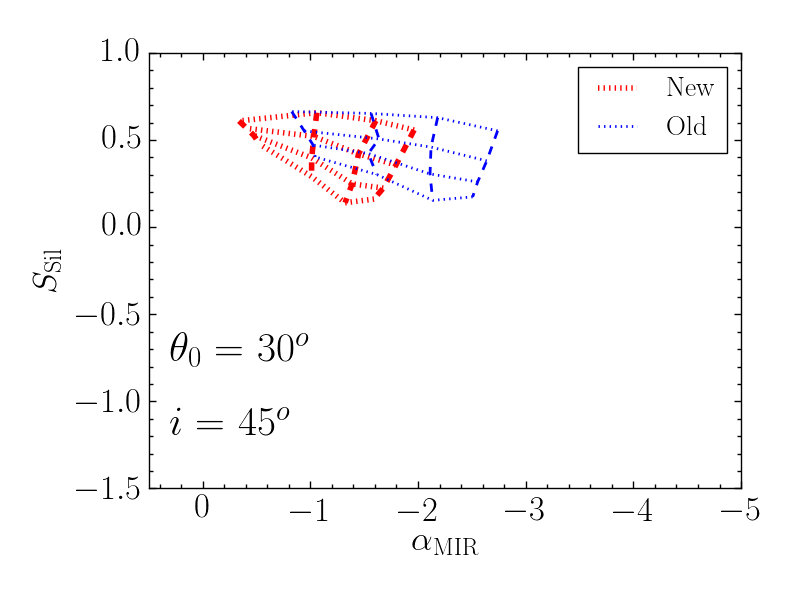}{\vspace{0cm}}
      \includegraphics[width=0.31\textwidth]{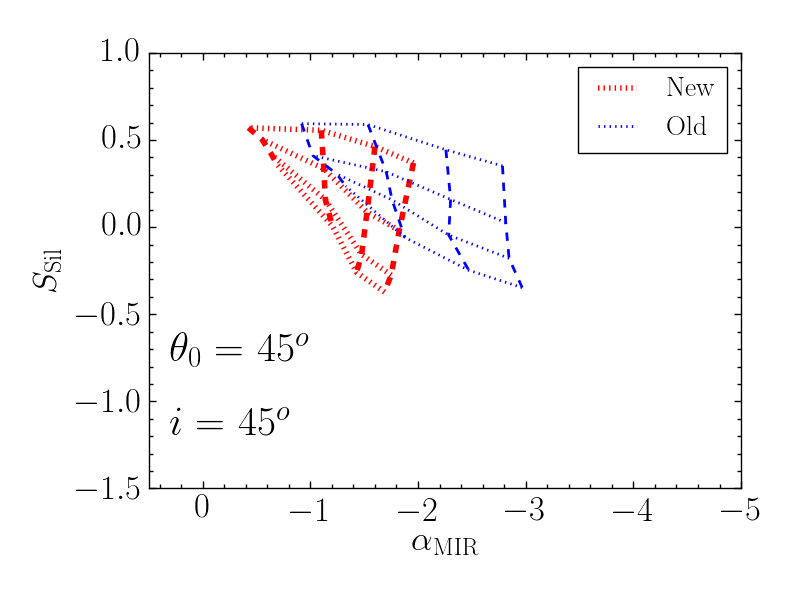}{\vspace{0cm}}
      \includegraphics[width=0.31\textwidth]{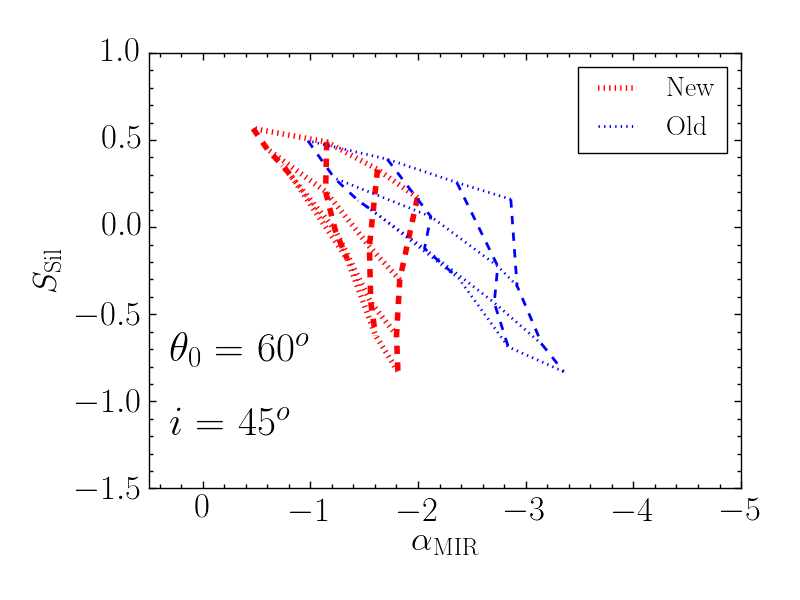}{\vspace{0cm}}    
      \includegraphics[width=0.31\textwidth]{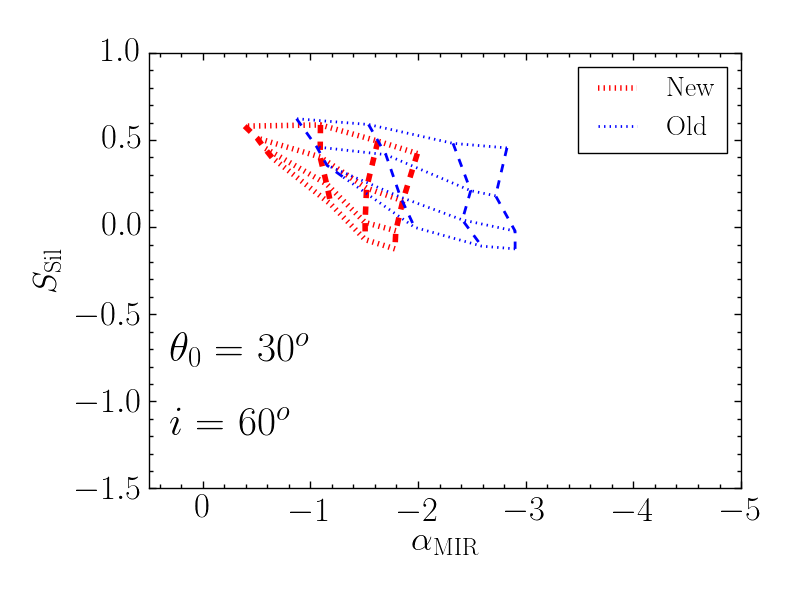}{\vspace{0cm}}
      \includegraphics[width=0.31\textwidth]{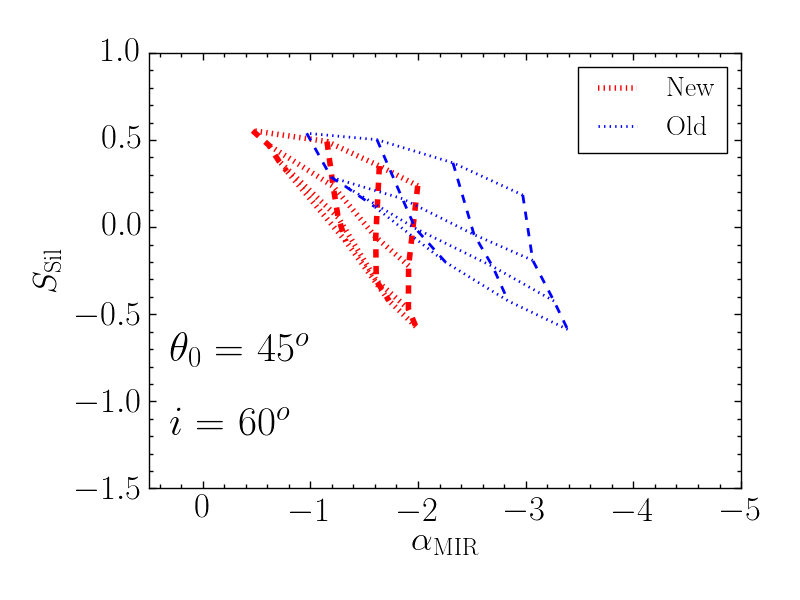}{\vspace{0cm}}
     \includegraphics[width=0.31\textwidth]{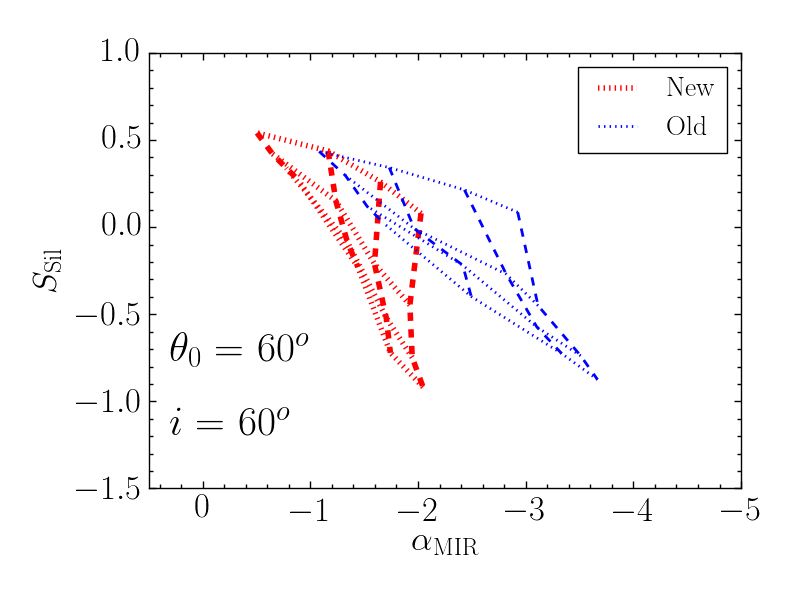}{\vspace{0cm}}    
      \includegraphics[width=0.31\textwidth]{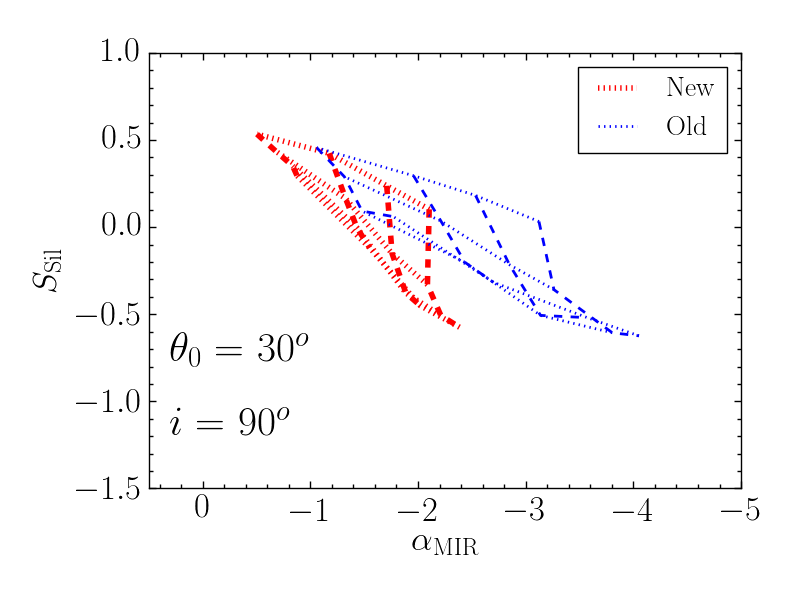}{\vspace{0cm}}
      \includegraphics[width=0.31\textwidth]{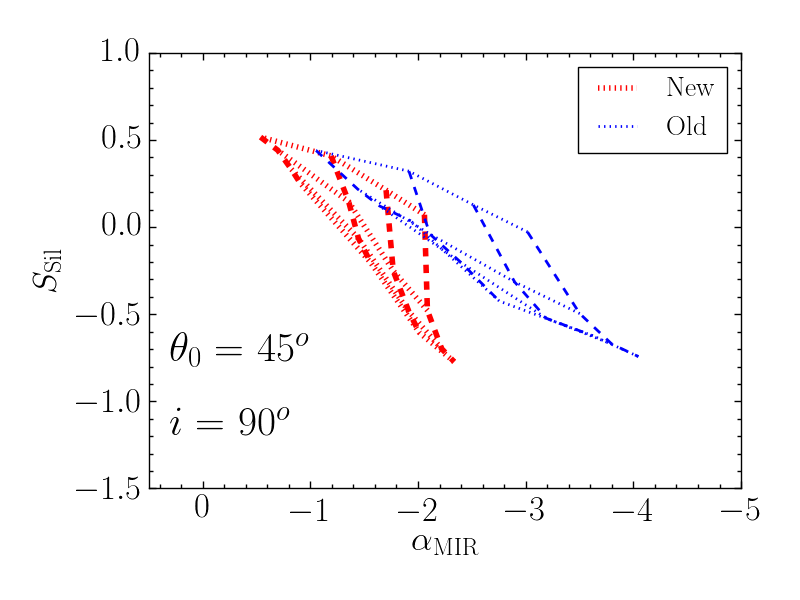}{\vspace{0cm}}
      \includegraphics[width=0.31\textwidth]{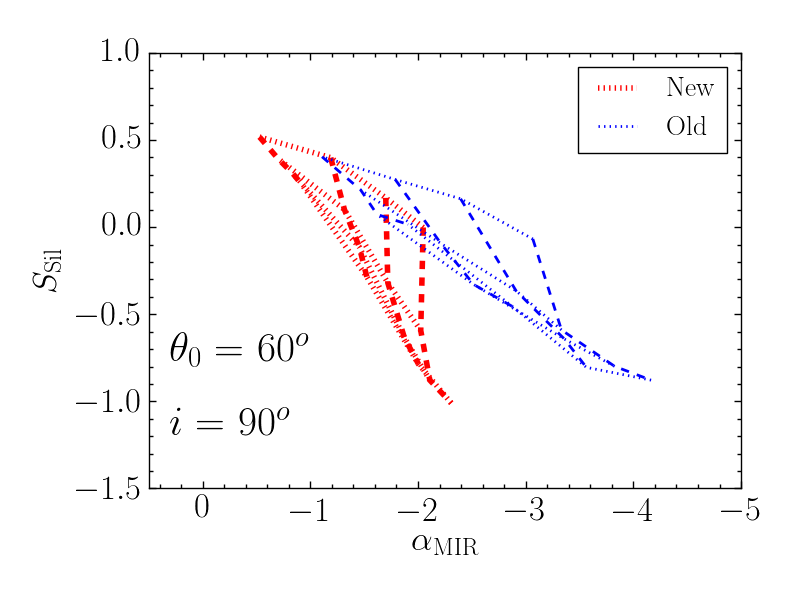}{\vspace{0cm}}

    \caption[Strength of the silicate feature against
      the MIR spectral index for the CAT3D models with the common parameters]{Comparison of the strength of the silicate feature and
    the MIR spectral index for the CAT3D old models (in blue),
    and the CAT3D new models (in red) only for the common parameters (see Table~\ref{tabla-parametros-modelos}). Each panel shows the estimated values for 
    different values (see top left panel) of $a$ (dotted lines) and $N_0$ (dashed lines) for a fixed inclination and
    torus half-covering angle. Results shown for five
    viewing angles, $i=0$\textdegree, 30\textdegree, 45\textdegree, 60\textdegree, and 90\textdegree.} 
   \label{figura-comparacion-modelos-same-parameters}

\end{figure*}
   


\clearpage

In Fig.~\ref{figura-comparacion-modelos} we show the calculated MIR spectral index against the strength of
the silicate feature for the old and new models. 
Each panel displays the  values for the entire range in
$a$ (dotted lines) and $N_0$ (dashed lines) for a fixed inclination and torus half-covering angle, $\theta_0$.
We only show results for five viewing angles ($i=0$\textdegree, 30\textdegree, 45\textdegree, 60\textdegree, and 90\textdegree),
one for each of the rows.  
As can be seen from these figures, for a given configuration of set
$\theta_0$ and $i$, fewer clouds along the equatorial direction tend to
produce weaker silicate features than configurations
with more clouds with a slight dependence with
the optical depth of the clouds   $\tau_{\rm V}$.
As explained in the previous section, increasing $N_0$ results in 
deeper silicate absorption features 
and reduced silicate emission features (see Section~\ref{dustsublimationmodel}).
 Interestingly, for
both the old and the new models for thin tori, $\theta_0$=30\textdegree,
and values of the inclination of $i\le 45$\textdegree,  the silicate feature is
always produced in emission. We can also see from these figures that
thicker tori tend to decrease the strength of the silicate feature when seen
in emission (that is, make it flatter) or make it deeper when the silicate is in absorption.

\begin{figure*}

      \includegraphics[width=0.45\textwidth]{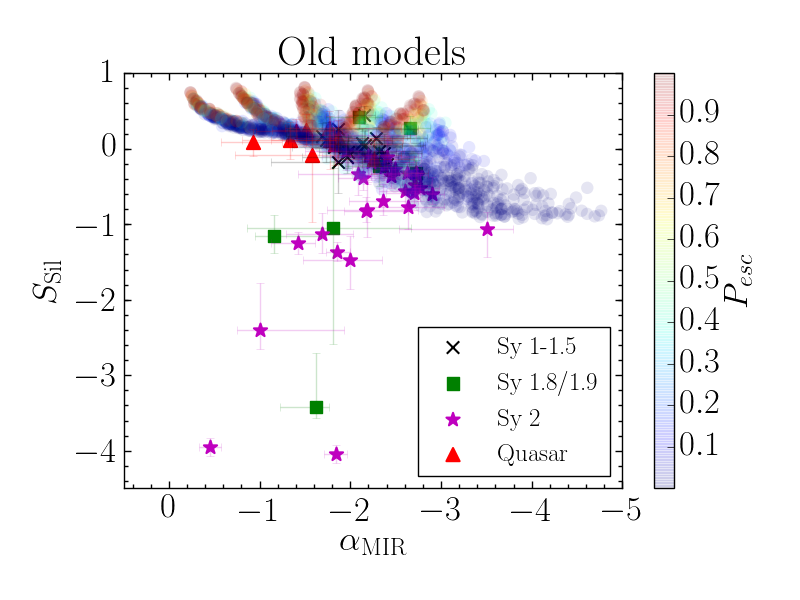}
      \includegraphics[width=0.45\textwidth]{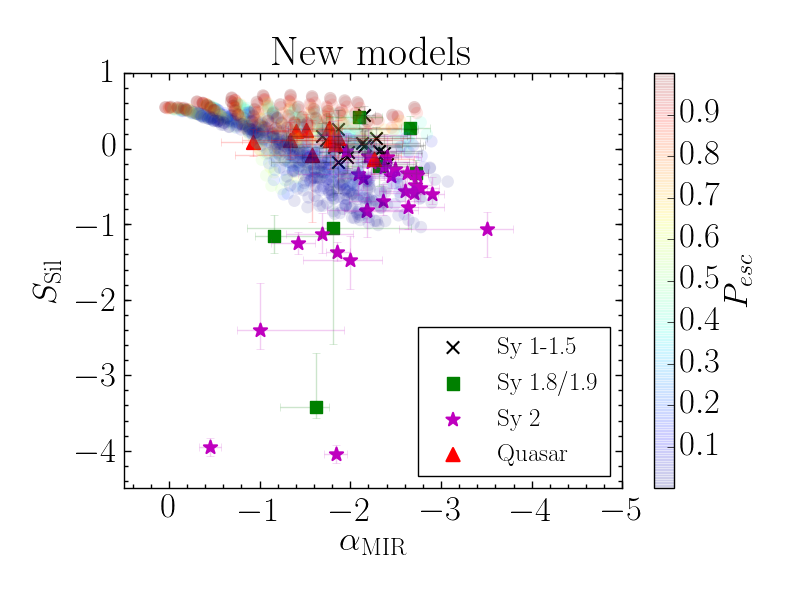}
      \includegraphics[width=0.45\textwidth]{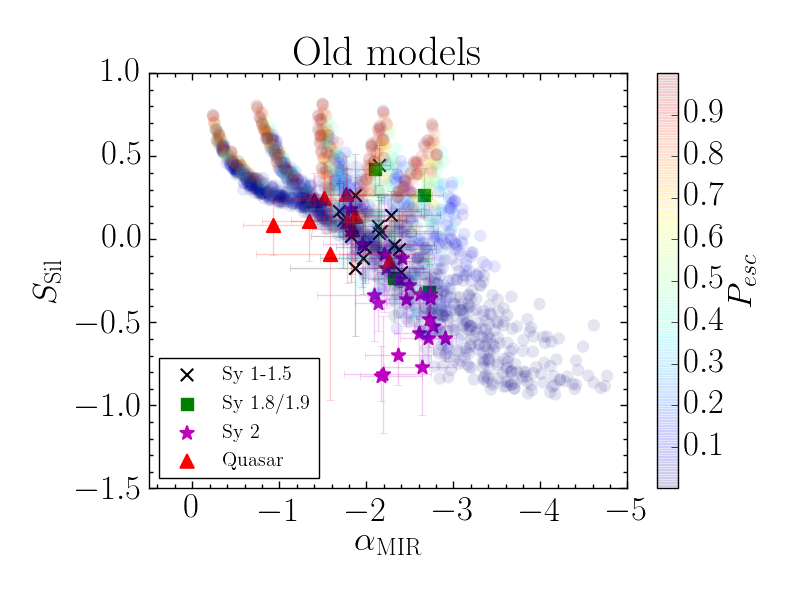}
      \includegraphics[width=0.45\textwidth]{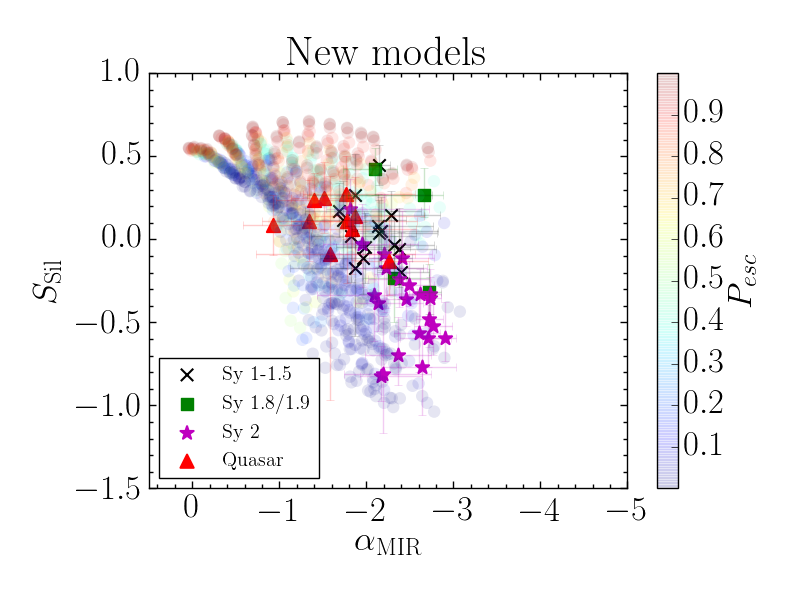}

     \caption{Top panels: MIR spectral index vs. the strength of the silicate
     the values for all parameters  (see Table~\ref{tabla-parametros-modelos}) 
     of the CAT3D  old torus models (left) and the new torus models (right).
     The model symbols (semi-transparent dots) are colour coded in terms of $P_{\rm esc}$  (see equation~\ref{eq:escape_probability}).
     The different types of Seyfert are shown as black crosses for 
   the Seyfert 1-1.5 nuclei, green squares for the Seyfert 1.8/1.9, magenta stars for the Seyfert 2 nuclei, and red triangles
for the   IR-weak quasars \citep{Alonso-Herrero2016b}.
   Bottom: Same as upper panels but excluding those galaxies not represented by the CAT3D models and
   thus zooming in the Y axis for $S_{\rm Sil}>-1.5$.  }
     \label{figura_modelos_todos_probabilidad_galaxias}

\end{figure*}

For the predicted MIR  ($8.1-12.5$\,$\mu$m) spectral index
there is a dependence with the index of the radial cloud distribution $a$
 and the half-covering angle of the torus $\theta_0$ in the
  sense that thicker tori and  flatter  dust radial cloud distributions
 (less negative values of the index of the radial distribution of the clouds
 $a$) produce more negative MIR spectral indices 
    (redder SEDs). This is because the flat and nearly flat distributions
  ($a=0.0$ and $a=-0.5$) have more cold dust at larger radial
  distances whereas for the steeper dust radial cloud distributions most of the dust is at small 
  radial distances from the AGN, so the average dust temperature is higher and have
  relatively more emission at shorter wavelengths. In the case of the  
  half-covering angle of the torus $\theta_0$, the effect is due to the self-obscuration
  within the dust distribution, i.e. for thicker tori we are 
  shielding more hotter clouds, making the overall SED redder.
 However, this strictly applies only to geometries and
 viewing angles were self-obscuration is not strong (that is, few clouds and low-intermediate
 values of $i$. On the other hand,
 $\alpha_{\rm MIR}$ for a given
 configuration of $\theta_0$, range of values of $a$ and  $\tau_{\rm V}$ has only
 a slight dependence on the viewing angle (i.e., compare models in the
 vertical panels of Fig.~\ref{figura-comparacion-modelos})
 with the spectral index becoming steeper
 for more inclined views.

 \cite{Hoenig-kishimoto}
  stated that there is very little dependence of model output SED on the assumed optical depth of the clouds and
  that  $\tau_{\rm V}=50$ for a standard ISM composition gives a good representation of observations.
  In Fig.~\ref{figura-comparacion-modelos} we can see indeed that the dependence of the MIR spectra
  l index and strength of the silicate feature on
  $\tau_{\rm V}$ is small for the old models. The only noticeable trend when the silicate feature is
  in emission is that the $\tau_{\rm V}=30$
  models always produce a stronger feature than the $\tau_{\rm V}=80$ models.  For inclinations
  $i>$45\textdegree \, when the feature is observed in absorption also the $\tau_{\rm V}=30$
  models always produce a stronger feature than the $\tau_{\rm V}=80$ models. 
  This is due to a  change in the MIR for the optical depth.
  This leads to source functions preferentially with deeper features and also produces deeper 
  features from self-absorption/extinction by other clouds.

In order to make a better comparison between the old and the new models
we repeat in Fig.~\ref{figura-comparacion-modelos-same-parameters}  the comparison of
the strength of the silicate feature and
the MIR spectral index for the old and new models but
only for the parameters in common. That is, $\tau_{\rm V} =$ 50;
$a = 0.00$, -0.50, -1.00, -1.50; and $N_0$ = 2.5, 5.0, 7.5, 10.0.
As for Fig.~\ref{figura-comparacion-modelos}, we only show five inclinations.
The most noticeable difference between the models is that for the same configuration
(same $i$, $\theta_0$, $a$ and $N_0$) the old models reach 
a more negative value of the MIR spectral index. This is expected due to the 
dust sublimation model introduced in the new models (Section~\ref{dustsublimationmodel}).
Due to self-obscuration effects becoming important further in, the silicate-bearing clouds are on average cooler, thereby contributing more to silicate absorption via obscuration/extinction than to emission.
For this reason, it is
necessary to include positive values of $a$ (that is, inverted radial distribution of clouds) for the new models,
in order to reach more negative values of $\alpha_{\rm MIR}$.
In the case of the strength of the silicate feature, 
the new models appear to produce always slightly deeper
silicate features with the differences becoming 
larger for the thickest tori and more inclined views.
For relatively inclined cases, this effect is due to the 
differential dust sublimation. The strongest emission features are always produced by 
the hottest silicate dust. While for the old models the hottest silicate temperature is 1500\,K,
for the new models it is 1250\,K. So the new models are starting out with a weaker silicate feature in emission.
When raytraced through the dust distribution, these get turned into absorption 
by self-obscuration/extinction. Since the new models started out with weaker features in emission
they end up with deeper absorption features.

\subsection{Comparison between old and new model predictions and observations}\label{sec:CAT3Dvsobs}

In this section we compare the predictions for the  MIR
emission of the old and new CAT3D models with our
observations of Seyfert galaxies and IR-weak quasars to see if the improved physical model
for the dust sublimation  produces a better description of the observations.
As the classification in Seyfert 1 or 2 is a probabilistic effect 
in torus models where the dust is distributed in clumps, to compare 
the models with the observations, we calculate for each model
the  probability that  an AGN produced photon escapes  unabsorbed using the 
following expression:

\begin{equation}
  \label{eq:escape_probability}
P_{\rm esc}=\exp\left(-N_0 \times exp \left( \frac{-(90-i)^2}{{\theta_0}^2}\right)\right)
\end{equation}

If the escape probability is high the models correspond statistically to a Seyfert 1 galaxy, and
if the probability is low to a Seyfert 2 galaxy \citep{Elitzur2012}. The idea is to have a gradual
transition between types instead of using an arbitrary value of the 
probability to separate the models between type 1 and  type 2.
We note that \cite{Hoenig2010} used an inclination criterion to separate
Seyfert 1 (inclination of $i=30$\textdegree) and Seyfert 2 (inclination of $i=75$\textdegree) models.
 However, \cite{Ramos-Almeida2011}
and \cite{AAH2011} used the clumpy torus models of \cite{Nenkova2008a,Nenkova2008b}
and a Bayesian approach to fit the IR SEDs of nearby Seyfert galaxies.
Both works demonstrated that the viewing angle is not the only
 determinant torus parameter to separate out Seyfert 1 and Seyfert 2, and
therefore the escape probability of an AGN-produced photon
is a better way to separate the Seyfert 1 models from the Seyfert 2 models \citep[see also][]{Elitzur2012}.

\begin{figure}
    \begin{center}
      
      \includegraphics[width=0.45\textwidth]{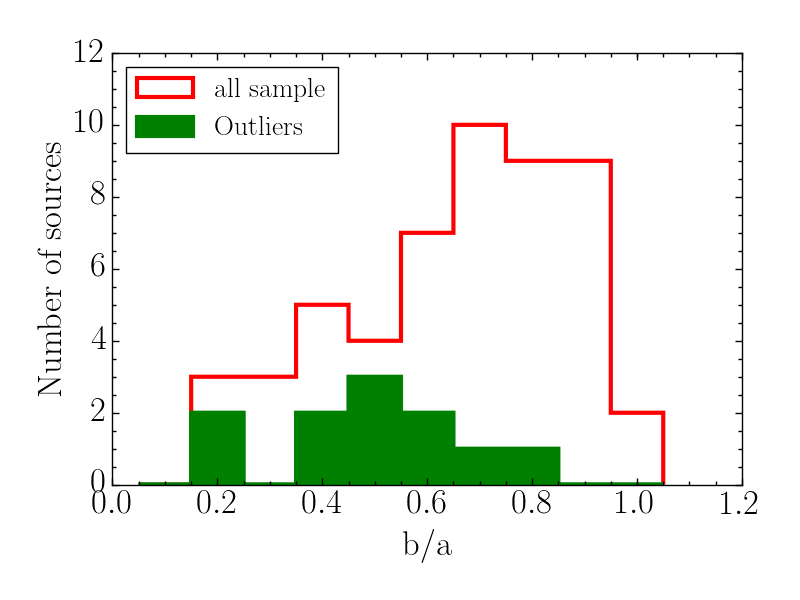}{\vspace{0cm}}
  
      \caption[b/a for all the galaxies and the outliers]{Distribution of the inclination of the host
        galaxies ($b/a$) for all the
        Seyfert galaxies (red histogram) and
  for the outliers, i.e., those Seyfert nuclei whose MIR properties are not represented by the CAT3D torus models (green filled histogram).}

  \label{fig_b_a_outliers}

  \end{center}
\end{figure}

Fig.~\ref{figura_modelos_todos_probabilidad_galaxias} compares the CAT3D model predictions 
 and the observations of the Seyferts and  quasars. 
Neither the new nor the old models explain
those galaxies with nuclear deep silicate absorptions, i.e., values of the 
strength of the silicate feature approximately $S_{\rm Sil}< -1$. This is similar to findings by other
works \citep{Levenson2007,Sirocky2008,AAH2011,GonzalezMartin2013} using the clumpy
torus models of \cite{Nenkova2008a,Nenkova2008b}.
 There are  11 galaxies in our sample whose values are not represented by the new  models and we
  study the possibility that they are objects with host obscuration.  Eight  of them are
classified as Seyfert 2 and 3 of them 
Seyfert 1.8/1.9 galaxies.  In Fig.~\ref{fig_b_a_outliers}
 we show the distribution of the inclination of the host galaxy ($b/a$) for all galaxies
and the galaxies that are not represented by the new models. 
We can see that the galaxies not represented by the CAT3D torus models
 tend to be in more edge-on galaxies (lower values of $b/a$) when compared with 
the entire sample. Other galaxies are in
  interacting systems or systems with disturbed morphologies
  (i.e., IC~4518W and NGC~7479).

A general result from Fig.~\ref{figura_modelos_todos_probabilidad_galaxias} is that the new CAT3D
models represent better the  distributions of the MIR spectral indices and strengths of
  the silicate features of the 
IR-weak quasars, the Seyfert 2 and Seyfert 1.8/1.9 galaxies than the old ones. The Seyfert 1-1.5 
are well represented with both the old and the new models.
It is also noteworthy that the old models for certain parameter configurations produce very steep
  MIR spectra indices (up to $\alpha_{\rm MIR}=-4$) for low escape probabilities that are not observed in
  Seyfert nuclei or local type 1 quasars.  
  We therefore conclude that the new models  with the 
improved dust physics reproduce better     the
  MIR properties of  the local type 1 (IR-weak) quasars and Seyfert galaxies.

  In Fig.~\ref{figura_modelos_todos_probabilidad_galaxias} the model predictions are
  colour coded according to $P_{\rm esc}$. The new
    CAT3D models have increased $P_{\rm esc}$ at the location of the type 1 AGN (Seyferts and quasars) whereas the old
    models had some difficulties producing relatively red MIR colours unless there were a
    lot of clouds (see next section) which makes $P_{\rm esc}$ low.
As expected, in this figure most
Seyfert 2 nuclei are close to models with low escape probabilities, although
the Seyfert 1-1.5 nuclei in this diagram are in a region populated by models with
both low and relatively high escape probabilities (see next section too).
However, for the new CAT3D models the majority of
silicate features in emission are observed for parameter configurations
resulting in relatively high escape  probabilities ($P_{\rm esc} >0.7$,
       approximately) as also found by \citet{Nikutta2009} for the {\sc CLUMPY} models.
  

\subsection{Constraining the CAT3D torus model parameters}

In the previous sections we compared the CAT3D old and  new torus models with all the Seyfert galaxies
and the IR-weak quasars.  In
this section we focus on the old and new models 
and  the Seyfert galaxies whose MIR properties are explained by the models.  The goal is to determine if we can constrain
some of the CAT3D clumpy torus model parameters from a statistical point of view. To do so,
for the observations we will use the combined PDFs of the Seyfert 1-1.5, Seyfert 2 (only those reproduced by
  the models, see Table~\ref{table-statistics-varias-figuras}) and  quasars \citep{Alonso-Herrero2016b}.
  
\begin{figure*}

      \includegraphics[width=0.45\textwidth]{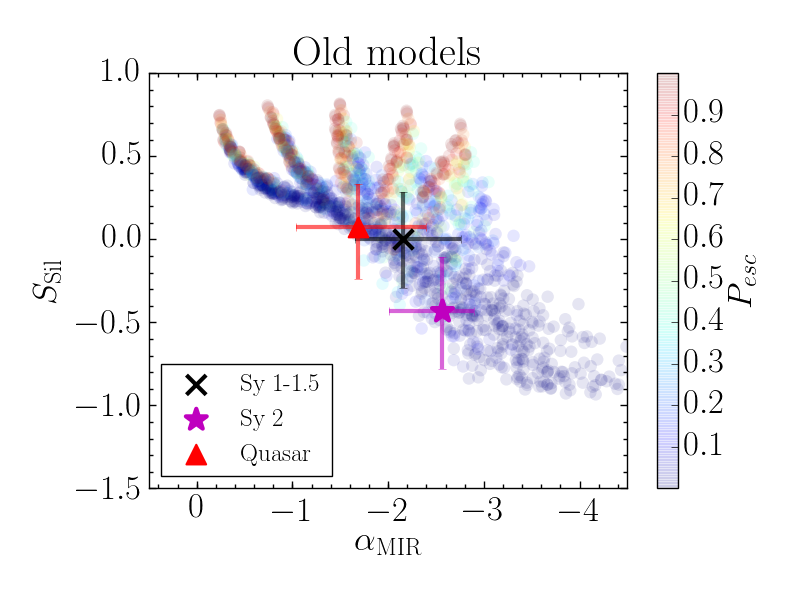}{\vspace{0cm}}
      \includegraphics[width=0.45\textwidth]{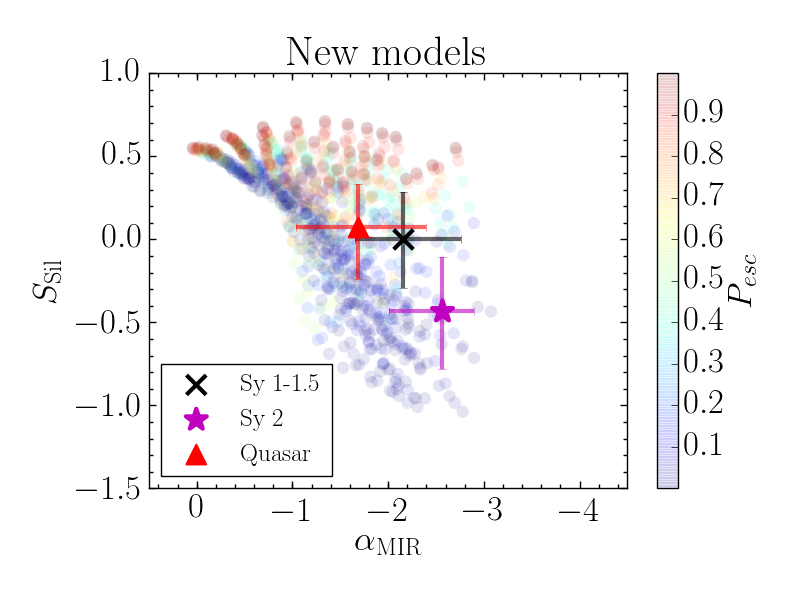}{\vspace{0cm}}
     
  
      \caption{Same as Fig.~\ref{figura_modelos_todos_probabilidad_galaxias} for the old
        and new model outputs. For the observations we plot the
        median values and
  $1\sigma$ uncertainties of the derived combined PDF of the  MIR spectral index and the strength of the silicate
        feature as a black cross for the Seyfert 1-1.5, a magenta star symbol for those Seyfert 2 galaxies reproduced by the models
        (see text) and a red triangle for the IR-weak quasars from \citet{Alonso-Herrero2016b}. }

  \label{fig_new_models_probability}

\end{figure*}

Figure~\ref{fig_new_models_probability} shows this comparison with the model symbols  colour coded in
     terms of $P_{\rm esc}$. As noted above, the Seyfert 2 galaxies are explained by models with
     low photon escape probability  for both the old and the new models. However, the Seyfert 1-1.5 galaxies and the quasars
      lie in a region of this diagram occupied by models with relatively high and low AGN photon escape probabilities.
     This is due to the degeneracy inherent to clumpy torus models, with models with
     different set of parameters producing the same values of the strength of 
     the silicates and the MIR spectral index.   On the other hand, detailed fits to the individual
       IR SEDs of Seyfert 1 with the clumpy torus
       models of \cite{Nenkova2008a,Nenkova2008b}  showed that the derived escape probabilities are never extremely
       high \citep[typically $P_{\rm esc}=0.2-0.3$, see][]{Ramos-Almeida2011,AAH2011,Audibert2017}. The escape
       probabilities of Seyfert 2 are generally found to be $P_{\rm esc}<0.1$. 
       This is also in good agreement with our statistical result that the models
       (both the old and the new ones)
       with very high photon escape probabilities ($P_{\rm esc} >0.7$,
       approximately) tend to occupy a region of the diagram not populated
       by the observations (i.e., approximately $\alpha_{\rm MIR}>-1$ and $S_{\rm Sil}>0.25$). 


\begin{figure*}

      \includegraphics[width=0.45\textwidth]{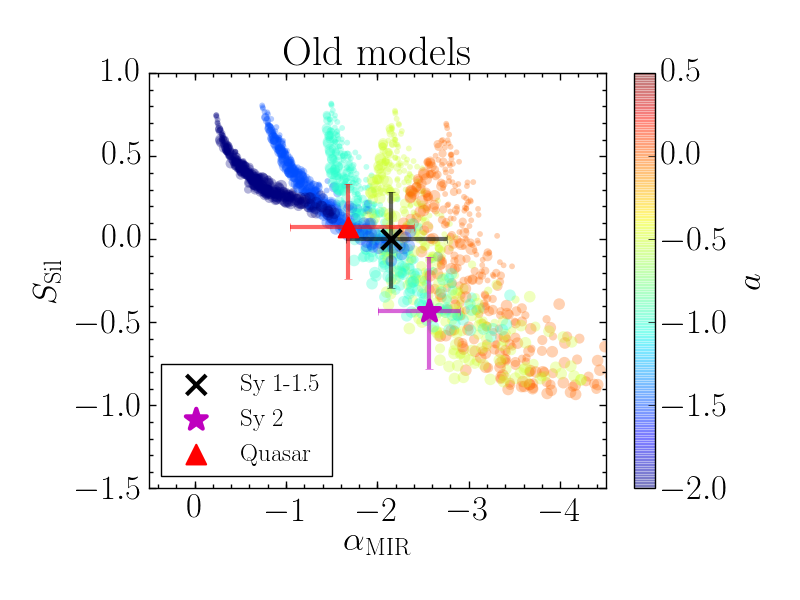}{\vspace{0cm}}
      \includegraphics[width=0.45\textwidth]{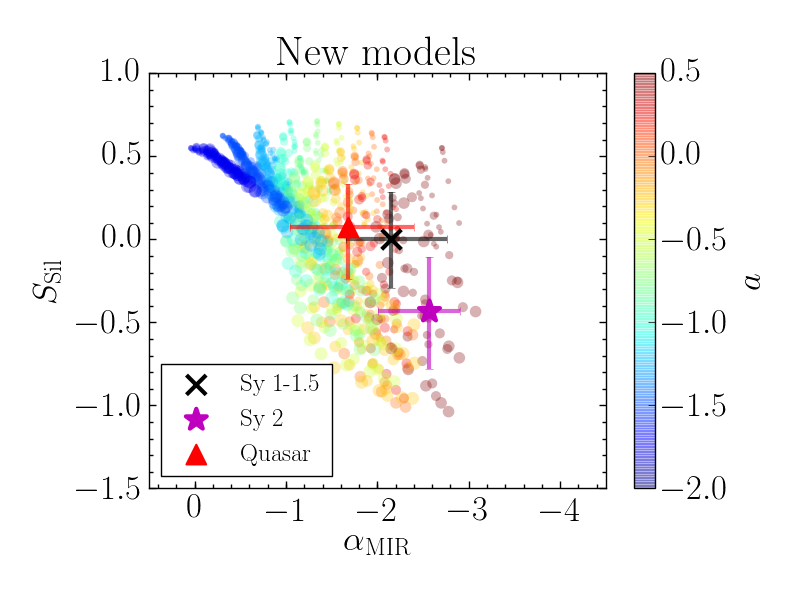}{\vspace{0cm}}
  
      \caption[As Fig.~\ref{fig_new_models_probability} but colour coded in terms of  $a$, and the size proportional 
        to $N_0$]{As Fig.~\ref{fig_new_models_probability} but the model symbols are colour coded in terms of the value the power-law index of 
          radial dust-cloud
     distribution, $a$, and the size of the model symbols is proportional 
        to the number of clouds along an equatorial line-of-sight,  with the smallest symbols
        corresponding to $N_0=2.5$ and the largest symbols to $N_0=12.5$ (see Table~\ref{tabla-parametros-modelos}).}

  \label{fig_new_models_a_N0}

\end{figure*}


\begin{figure*}
      
      \includegraphics[width=0.45\textwidth]{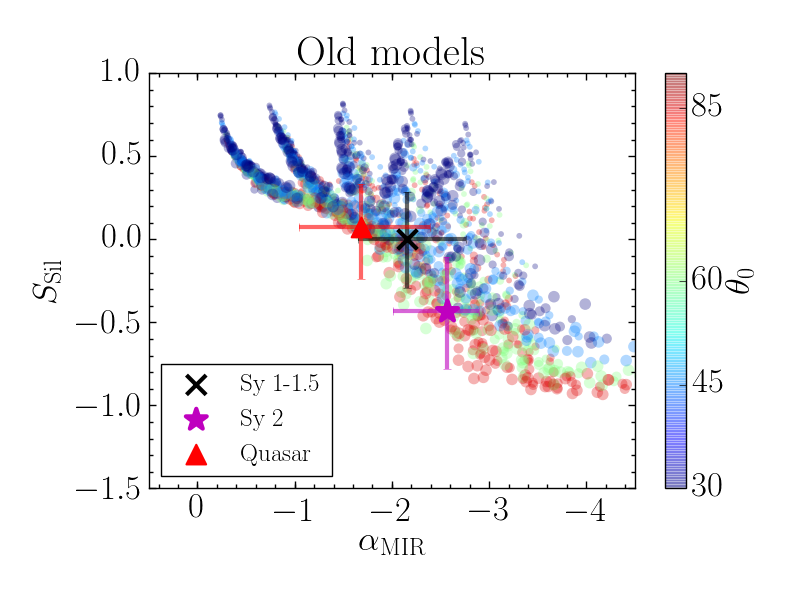}{\vspace{0cm}}
      \includegraphics[width=0.45\textwidth]{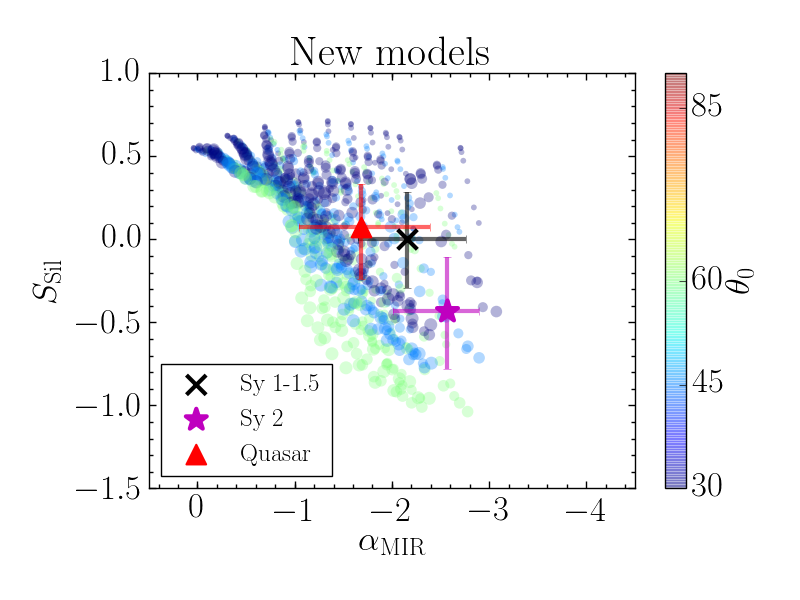}{\vspace{0cm}}
  
      \caption[As Fig.~\ref{fig_new_models_probability} but colour coded in terms of $\theta_0$, and the size proportional 
        to $N_0$]{As Fig.~\ref{fig_new_models_probability} but the model symbols are colour coded
in terms of the value the torus half-covering angle, $\theta_0$, 
     and the size of the model symbols is proportional 
     to the number of clouds along an equatorial line-of-sight, $N_0$ as in Fig.~\ref{fig_new_models_a_N0}.}

  \label{fig_new_models_theta_N0}

\end{figure*}


\begin{figure}
      
      \includegraphics[width=0.45\textwidth]{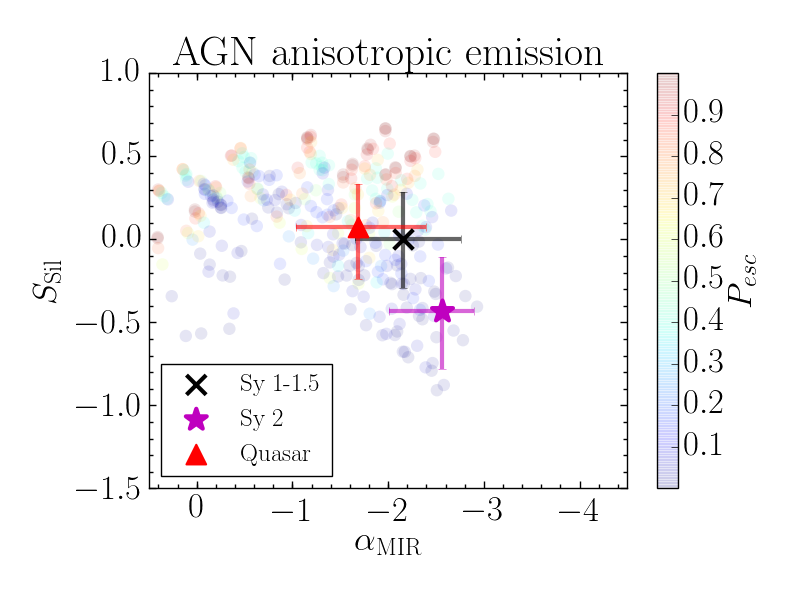}{\vspace{0cm}}
      \includegraphics[width=0.45\textwidth]{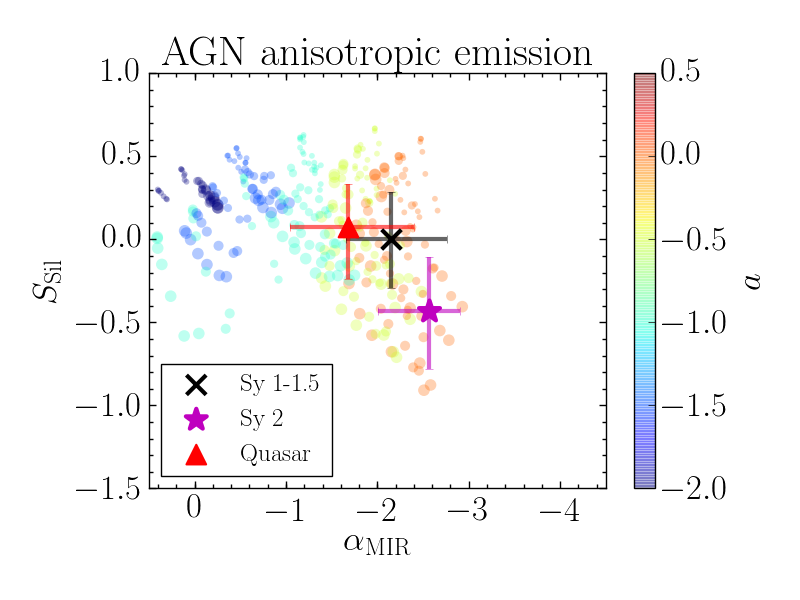}{\vspace{0cm}}
      \includegraphics[width=0.45\textwidth]{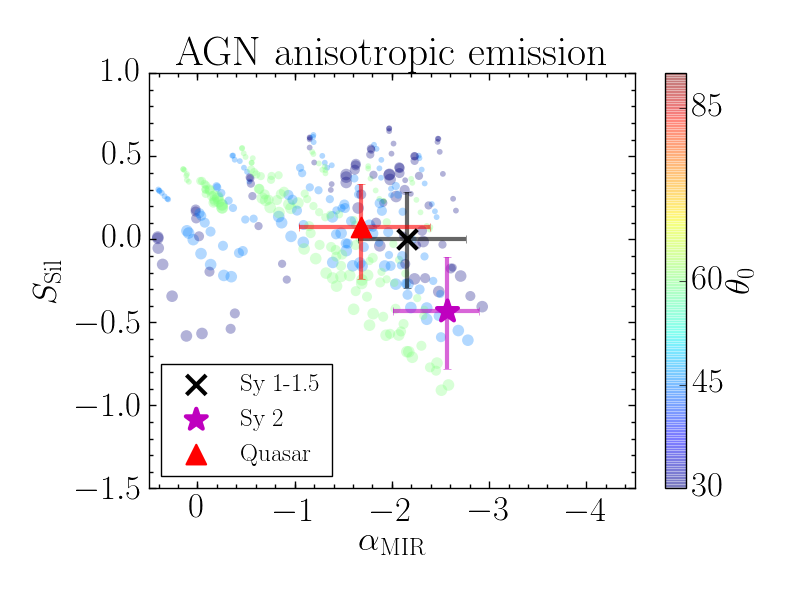}{\vspace{0cm}}
  
      \caption{As the right panels  of  Figs.~\ref{fig_new_models_probability} (top),~\ref{fig_new_models_a_N0}
     (middle) and \ref{fig_new_models_theta_N0} (bottom)
but showing the results for the new CAT3D  models with anisotropic AGN emission.}

  \label{fig_new_models_anisotropic}

\end{figure}


In Fig.~\ref{fig_new_models_a_N0} we show the same comparison 
 as in Fig.~\ref{fig_new_models_probability}, but now  we colour code
the model symbols in terms of the value of the power-law index of the radial dust-cloud
distribution, $a$. The size of the model symbols is proportional 
to the number of clouds along an equatorial line-of-sight, $N_0$.
From a statistical point of view Seyfert 2 nuclei
are reproduced with models with  more clouds in the 
  equatorial direction than type 1 AGN.   For the new models, those Seyfert 1-1.5 galaxies
  in our sample with a nearly flat
  silicate feature ($S_{\rm Sil}\sim0$) can
also be reproduced with models with more clouds, whereas the Seyfert 1-1.5 galaxies
with silicate emission 
are only reproduced with a few clouds in the equatorial direction
 (see also Fig.~\ref{figura_modelos_todos_probabilidad_galaxias}
  where we plotted the values of the individual
  objects). As explained in previous sections, the increase in the number
  of clouds along the equatorial line-of-sight, $N_0$, reduces the strength of the silicate
  feature in emission
  or turns them into absorption features.
  This is in good agreement with the conclusions of \cite{Hoenig-kishimoto}, who  showed
that more clouds along an equatorial direction direction (larger values of $N_0$) produce weaker
emission at 10\,$\mu$m in Seyfert 1.  For Seyfert 2, the  silicate feature is deeper for
large values of $N_0$.
The same result  was obtained by \cite{Ramos-Almeida2014b} with the 
clumpy torus models of \cite{Nenkova2008a,Nenkova2008b}. They found that for Seyfert 1,
flat silicate features can also be reproduced with high values of $N_0$ ($\sim 10-15$)
  and high values of the optical depth $\tau_V=100-150$,
whereas strong silicates in emission are produced by configurations with a few optically
thin clouds along the equatorial direction. For Seyfert 2,
the silicate feature in absorption is also reproduced with high $N_0$ ($\sim 8-15$)   with $\tau_V \sim 50$. 
\cite{Ichikawa2015} also used the \cite{Nenkova2008a,Nenkova2008b} clumpy models
and found that there were statistically significant differences between the distributions of
$N_0$ for the Seyfert 1 and the Seyfert 2 with hidden broad line region.
For the old models there is more degeneracy in
terms of the number of clouds for the Seyfert 1-1.5 galaxies and they can be explained 
with models with low and high number of clouds.

From Fig.~\ref{fig_new_models_a_N0} we can also set a limit on $a$ values that 
can reproduce our Seyfert galaxies using the CAT3D models.
 Very negative values of $a$, i.e., $a \leq -2.0$ for 
the old models and  $a \leq -1.5$ for the new models, 
cannot reproduce the values observed in Seyfert galaxies or even the  quasars.
To represent the Seyfert 2 values it is necessary to have
positive values of $a$ ($a=0.25, 0.50$), that is, radial distributions with more clouds towards the outer parts
of the torus.
  We also found steeper radial distributions of clouds in the 
 old models than in the new ones. We can also conclude
 that there is a tendency for  quasars, Seyfert 1-1.5 and Seyfert 2 to be
   reproduced with increasingly flatter indices of radial distributions of the torus clouds (more positive
   values of $a$). This is in good agreement with the result of \citet{Martinez-Paredes2017} using a detailed
   modelling of the nuclear SEDs with the clumpy torus models of \citet{Nenkova2008b}.

 We finally compare the data and the models in 
  Fig.~\ref{fig_new_models_theta_N0} colour coding 
the model symbols in terms of the torus half-covering angle, $\theta_0$. 
 We can observe  a tendency for the Seyfert 1-1.5 and the IR-weak  quasars to 
be represented with relatively thinner tori ($\theta_0\le 45$\textdegree)
than the Seyfert 2 galaxies  (old models), but there is a degeneracy produced by the clumpy torus models.
This is consistent with the finding of thinner tori  in Seyfert 1-1.5 than Seyfert 2
  by \cite{Ramos-Almeida2011} and \cite{Ichikawa2015} using 
 fits of the IR SEDs of Seyfert nuclei. For the new models, this tendency is not observed and all galaxies
are better represented with relatively thinner tori.
To break this degeneracy, it is required to have information about
  the nuclear near-IR emission \citep{Ramos-Almeida2014b}.

 Summarizing,  we are able to constrain
some of  the parameter ranges of the old and new CAT3D torus models. In particular, we set a lower limit to the index of the power-law
radial distribution of clouds  ($a\ge -1.5$ for the old models and $a\ge -1.25$ for the new models). We derive
statistical tendencies for type 1 nuclei, which are represented better with steeper dust radial distributions and thinner tori
than Seyfert 2 for the old models, whereas there is more degeneracy for the new models.
We also find that the MIR properties of Seyfert 2 nuclei are well reproduced
with CAT3D models with a combination of parameters that result in smaller escape
probabilities of AGN-produced photons.

\subsection{Models with AGN anisotropic emission}
\label{sec:anisotropicAGN}
We finally explore very briefly the effects on the MIR properties of introducing AGN anisotropic emission in the
new CAT3D models to take into account the expected angular dependence of an AGN UV emission.
We introduced a $\cos(i)$ dependence in the AGN illumination of the torus clouds as an approximation of the
more general angular dependence of the UV
radiation of an accretion disk $\propto 1/3 \, \cos(i)*(1+2\,\cos(i))$ \citep[see][and references therein]{Netzer1987}, also
adopted by other works \citep[e.g.,][]{Hoenig2006,Schartmann2005,Schartmann2008,Stalevski2012}.
 We run models with torus parameters identical 
to those listed in Table~\ref{tabla-parametros-modelos}
for the new models except for
the power-law index of the radial distribution of the clouds, which is in the $a=[-2,0]$ range and in steps of 0.5. 
The range is the same as in the old models in order to avoid the inverted radial cloud distributions
($a$ > 0).

The anisotropy introduces a significant additional vertical temperature gradient on top of self-obscuration.
For isotropic AGN emission, the projected radial temperature profile of the torus does not depend strongly
on $\theta_0$ (or the scale height of the torus) since the temperature is approximately the same on the  {\it skin} of the
torus. However, in the anisotropic case, the torus surface has a
lower temperature for a given
distance from the AGN when the scale height is small since the incident radiation from the AGN is adjusted
by $\cos(i)$. This results in redder SEDs (that is, steeper MIR spectral
indices, see Fig.~\ref{fig_new_models_anisotropic}). We note however, that \citet{Stalevski2012} found no
significant differences in the shape of the SEDs when including the AGN anisotropic emission in their
two-phase media clumpy
models. The difference in MIR predictions for CAT3D models with AGN isotropic and anisotropic emission 
is clearly seen in terms of the index of the radial distribution $a$. Models
with indices ranging from  $a=-2.0$ to $a=0$ and AGN anisotropic  emission reproduce the observed range of $\alpha_{\rm MIR}$ values
for Seyferts and quasars without the need for an inverted (i.e., $a>0$) radial
distribution of clouds. We also note that these new models would be able to
explain the rather flat MIR SEDs and shallow silicate absorptions
of low luminosity AGN \citep{Gonzalez-Martin2015}. However, as can also be seen from the comparison of  the right panel of
Fig.~\ref{fig_new_models_a_N0} and middle panel  of Fig.~\ref{fig_new_models_anisotropic},
adding AGN anisotropic  radiation alters the rather well-behaved dependence
of the MIR spectral index with the index of the cloud radial distribution.

Finally, we can see that AGN anisotropy has a small effect on the
  strength of the silicate feature \citep[compare e.g.,  right panel of Fig.~\ref{fig_new_models_theta_N0} and
  bottom panel of Fig.~\ref{fig_new_models_anisotropic},
    and see also][]{Schartmann2005,Stalevski2012}. Moreover, anisotropy does not suppress the $9.7\,\mu$m silicate feature in emission, as found
  by \cite{Mankse1998} for a flared disk model with AGN anisotropic emission.

\section{Summary and conclusions}
\label{conclusions}

The main  goal of this work was to make a statistical comparison of the MIR  properties of Seyfert nuclei and predictions
from the CAT3D clumpy torus models of \cite{Hoenig-kishimoto}.  We used the published 
version of the models (old models) with a standard ISM
dust composition. We also presented new calculations 
including an improved physical representation of the dust sublimation properties (new models) and AGN anisotropic emission.
The new CAT3D models allow graphite grains to persist at temperatures higher than the silicate dust sublimation
  temperature. This produces bluer NIR-to-MIR SEDs and flatter MIR spectral slopes.

We compiled ground-based MIR ($\sim 7.5-13.5\,\mu$m) spectroscopy of 52 nearby (median distance of 36\,Mpc) Seyfert galaxies, using
published observations taken with $8-10\,$m class telescopes with sub-arcsecond angular resolution.
The sample contains fifteen Seyfert 1-1.5, seven Seyfert 1.8-1.9, and thirty Seyfert 2. They are
located at a median distance of 36 Mpc and the  ground-based slits cover typical nuclear regions of 101 pc  in size.
We decomposed the spectra using \textsc{deblend}IRS to disentangle the AGN MIR emission from the  stellar and the PAH emission
arising from the  host galaxy and focused on the derived AGN MIR spectral index ($\alpha_{\rm MIR}$)
and strength of the $9.7\,\mu$m silicate feature ($S_{\rm Sil}$).



The CAT3D models do not reproduce the Seyfert
galaxies with deep ($S_{\rm Sil} < -1$) silicate absorption (approximately 20\% of the sample), as found with other clumpy torus
models. These galaxies tend to
have low values of $b/a$ (highly inclined galaxies) and some of them are  mergers. These are likely objects with contamination
from  obscuration  in the host galaxy \citep{AAH2011,Goulding2012,GonzalezMartin2013}.  Excluding these galaxies,
the new CAT3D models  improve overall the
representation  of the quasars, Seyfert 1.8/1.9  and Seyfert 2 galaxies. 

We also attempted to constrain the new CAT3D torus model parameters from a statistical point of view using the
MIR observations. The Seyfert 2 galaxies are well reproduced with low photon escape probability models, as expected, whereas
the type 1 AGN tend to have higher escape probabilities. The moderate silicate
features in absorption of Seyfert 2 are reproduced with models with more clouds along an equatorial
direction ($N_0$),
whereas the Seyfert 1-1.5 galaxies and the IR-weak quasars with silicate emission are explained with a few clouds.
There is also a tendency from quasars to Seyfert 1-1.5 to Seyfert 2 nuclei  to show increasingly shallower radial cloud
distributions (less negative values of $a$).
This is in good agreement with
previous works \citep{Hoenig-kishimoto, Ramos-Almeida2011,Ichikawa2015,Martinez-Paredes2017}.
Very negative values of
 $a$, i.e., $a \leq -2.0$ for the new models (that is, most of the
clouds are concentrated towards the inner regions of the torus)
tend to produce flatter MIR spectral indices and stronger silicate features
than observed in Seyfert galaxies or even the quasars.
In the new models most Seyfert 2 galaxies would require inverted radial cloud distributions (positive $a$).
  The problem with this uncommon geometry is solved
  by introducing a $\cos i$ dependency on the AGN illumination of the clouds (AGN anisotropic emission) which produces
  bluer MIR spectral indices in good agreement with the range of observed values.
%

In conclusion, including a more realistic dust
  sublimation physics as well as anisotropic AGN emission in the
new CAT3D models
reproduces better the overall MIR properties of local 
Seyfert 1-1.5, 2, and quasars. However, we cannot break fully the  degeneracy in all parameters of the CAT3D models (or any other
  clumpy torus models) by using MIR spectroscopy alone \citep[see also][]{Ramos-Almeida2014b} 
  even after isolating the AGN component. However, by using a large sample
  of Seyfert galaxies we were able uncover some
  differing trends between type 1-1.5 and type 2 in terms of the 
  index of the radial distribution of the clouds $a$ and the number of clouds 
  along the equatorial direction $N_0$.

  \section*{Acknowledgements}
 We thank the referee for valuable comments that helped improve the paper.
We acknowledge
financial support from the Spanish Ministry of Economy and Competitiveness
through the Plan Nacional de Astronom\'{i}a y Astrof\'{\i}sica  under grant AYA2015-64346-
C2-1-P (JG-G, AA-H), which was partly funded by the FEDER programme.
AH-C acknowledges funding by the Spanish Ministry of Economy and Competitiveness
under grant AYA2015-63650-P.
JG-G acknowledges financial support from the Universidad de Cantabria through the Programa de 
Personal Investigador en Formaci\'on Predoctoral de la Universidad de Cantabria.
AA-H is also partly funded by CSIC/PIE grant 201650E036. 
SFH acknowledges support from the Marie Curie International Incoming Fellowship within the
Seventh European Community Framework Programme (PIIF-GA-2013-623804) and the European Research
Council under Horizon 2020 grant ERC-2015-StG-677117.
CRA acknowledges the Ram\'on y Cajal Program of the Spanish Ministry of Economy and Competitiveness through
project RYC-2014-15779 and the Spanish Plan Nacional de Astronom\'{\i}a y Astrofis\'{\i}ica under grant AYA2016-76682-C3-2-P.
O.G.-M. acknowledges support from PAPIIT IA100516. C.P. acknowledges support from the NSF-grant number 1616828.
MK acknowledges support from JSPS under grant 16H05731.

Based on observations made with the GTC, installed in the Spanish
Observatorio del Roque de los Muchachos of the Instituto de
Astrof\'{\i}sica de Canarias, in the island of La Palma.
Based on observations obtained at the Gemini Observatory, which is operated by the Association
of Universities for Research in Astronomy, Inc., under a cooperative agreement with the NSF on
behalf of the Gemini partnership: the National Science Foundation (United States), the National Research Council (Canada), CONICYT
(Chile), Ministerio de Ciencia, Tecnolog\'{i}a e Innovaci\'{o}n Productiva (Argentina), and
Minist\'{e}rio da Ci\^{e}ncia, Tecnologia e Inova\c{c}\~{a}o (Brazil). 
Based on observations made with ESO Telescopes at the La Silla Paranal Observatory.
This research has
made use of the NED which is operated by JPL, Caltech, under contract
with the National Aeronautics and Space Administration. This
work is based in part on observations made with the Spitzer Space
Telescope, which is operated by the Jet Propulsion Laboratory, California
Institute of Technology under a contract with NASA.

\bibliographystyle{mnras}
\bibliography{references} 

\begin{figure*}
    \begin{center}
      
      \includegraphics[width=0.45\textwidth]{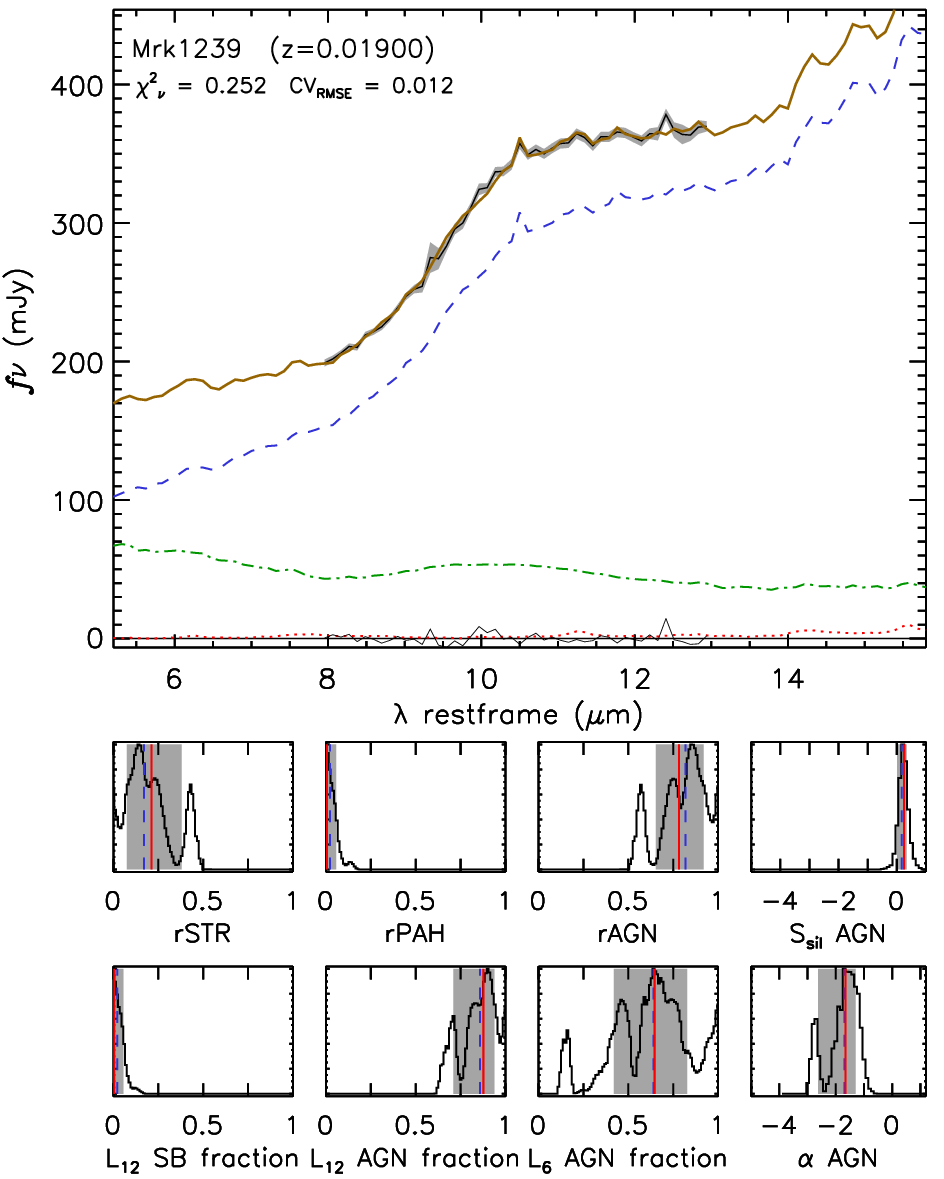}{\vspace{0cm}}
      \includegraphics[width=0.45\textwidth]{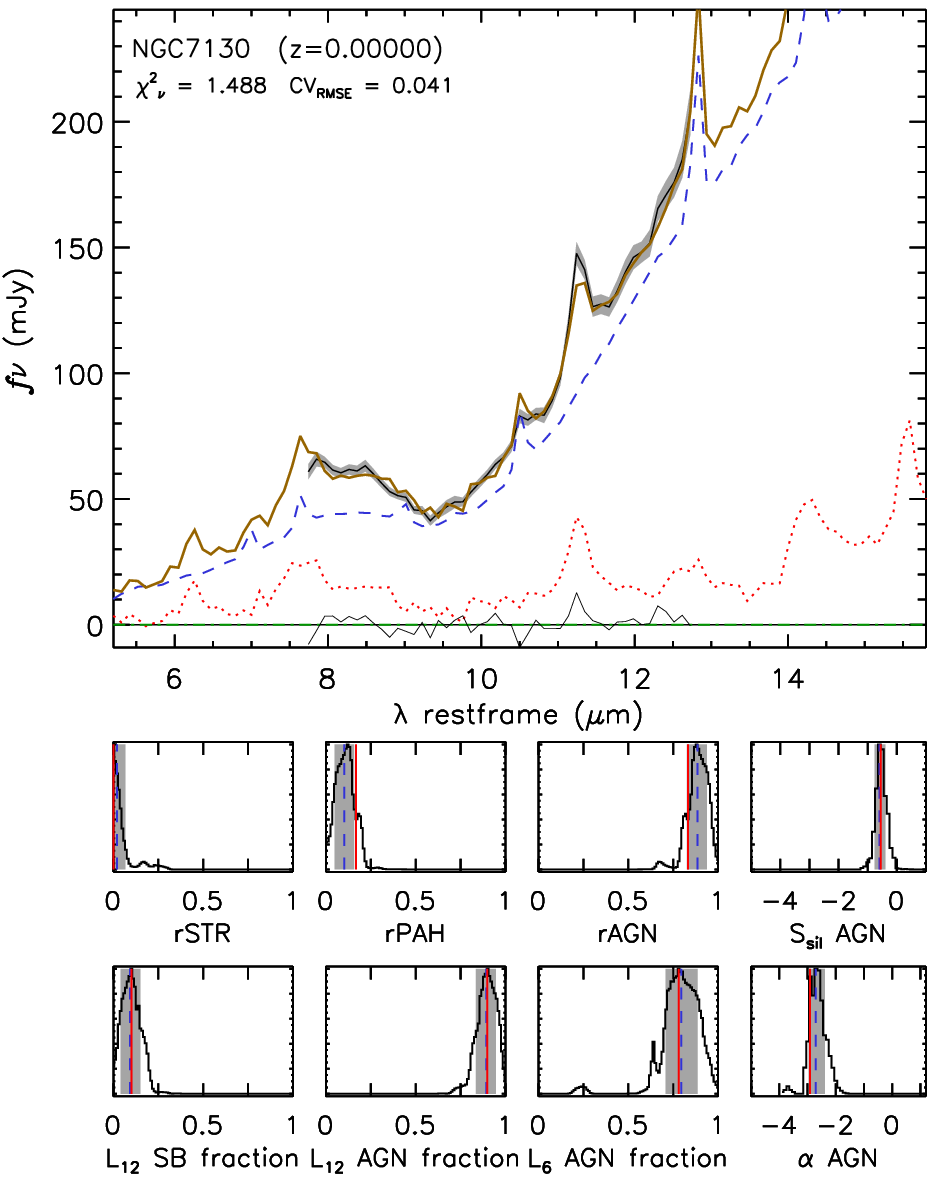}{\vspace{0cm}}

      \caption[Examples of the output of \textsc{deblend}IRS]
      {Examples of the output of \textsc{deblend}IRS for Mrk~1239 (Seyfert 1-1.5, left panel),
      and  NGC~7130 (Seyfert 2, right  panel).  For each example, the top panels show the rest-frame spectrum 
      with the best fitting model (orange), and the three components: stellar (dash-dotted green), PAH (dotted red)
      and AGN (dashed blue). In the bottom panels \textsc{deblend}IRS
      shows the PDF for the STR, PAH and AGN emission fraction within
      the slit ($5-15$\,$\mu$m), 
      namely rSTR, rPAH, and rAGN, respectively; the strength of the $9.7\,\mu$m
      silicate feature ($S_{\rm Sil}$)  and the spectral index ($\alpha$, or $\alpha_{\rm MIR}$ in this work notation) 
      in the AGN  spectrum ($8.1-12.5\,\mu$m); the fractional contribution
      within the slit of the AGN to the rest-frame 6\,$\mu$m ($L6_{\rm AGN}$) and  
      12\,$\mu$m ($L12_{\rm AGN}$) luminosity; and the  fractional contribution of the host galaxy to the rest-frame 12\,$\mu$m luminosity ($L12_{\rm SB}$).
      For the PDFs the solid red line indicates the value for the best fitting decomposition model whereas the dashed blue line
      indicates the expectation value. The  shaded area represents the 16\% and 84\% percentiles, i.e., the 1$\sigma$ confidence interval.
} 
   
  \label{fig_ejemplos_deblendIRS}

  \end{center}
\end{figure*}

\section*{Appendix}

The main output of \textsc{deblend}IRS is the best-fit combination of the stellar, interstellar and AGN components.
It also calculates the 
probability distribution functions (PDF) of the 8 parameters shown in Fig.~\ref{fig_ejemplos_deblendIRS} where we show an
example of 
a Seyfert 1.5
(Mrk~1239) and and example of a Seyfert 2 galaxy
(NGC~7130). We refer the reader to \citet{Antonio2015} for a full description of \textsc{deblend}IRS. 

\begin{figure*}

      \includegraphics[width=0.45\textwidth]{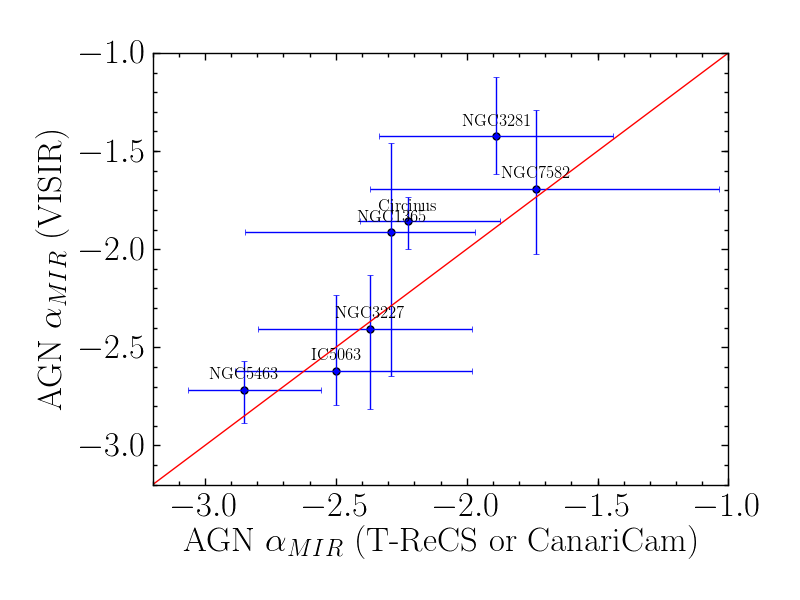}{\vspace{0cm}}
      \includegraphics[width=0.45\textwidth]{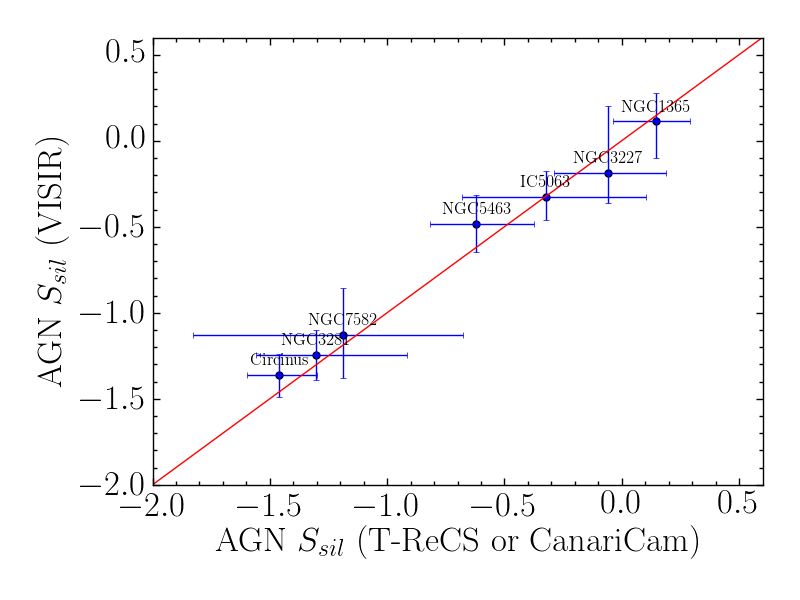}{\vspace{0cm}}

      \caption[Comparison of \textsc{deblend}IRS output for galaxies observed with two different instruments]
      {Comparison of the values obtained with \textsc{deblend}IRS for the 
      AGN MIR spectral index (AGN $\alpha_{\rm MIR}$, left panel) and the silicate strength (AGN $S_{\rm Sil}$, right panel)
      for the galaxies observed with two different instruments. The error bars 
      represent the 1$\sigma$ confidence interval (16\% and 84\% percentiles). The solid
      red line is not a fit but it represents the 1:1 relation.} 
   
  \label{fig_dos_espectros_por_galaxia}

\end{figure*}

For the galaxies observed with two instruments, we performed the 
\textsc{deblend}IRS decomposition for each one of the spectra.
In order to select the best one, we compared the values obtained 
for the AGN spectral index and the AGN 
silicate strength. For each galaxy we selected the 
spectrum for which estimated $\alpha_{\rm MIR}$ and $S_{\rm Sil}$
had the smallest 1$\sigma$ confidence interval (16\% and 84\% percentiles), i.e.,
the one with smaller error bars in Fig.~\ref{fig_dos_espectros_por_galaxia}. These in turn are the spectra
  with the lowest estimated errors.
 The observed spectrum corresponding to the best-fit is marked in bold in Table~\ref{table-instruments}.
From Fig.~\ref{fig_dos_espectros_por_galaxia} we can see that for each galaxy the 
silicate strengths fitted from the different instrument spectra are similar.
In the case of the AGN spectral index, for each galaxy the values from the 
two spectra are mostly compatible  within the 1$\sigma$ confidence interval.


\bsp	
\label{lastpage}
\end{document}